\documentclass[aps,prd,showkeys,superscriptaddress,singlecolumn,nofootinbib,floatfix]{revtex4-2}

\usepackage{subfiles}
\usepackage{amsmath}
\usepackage{amssymb}
\usepackage{amsthm}
\usepackage{mathrsfs}
\usepackage{graphicx}
\usepackage{epstopdf}
\usepackage{fancyhdr}
\usepackage{array}
\usepackage[all]{xy}
\usepackage{eufrak}
\usepackage{euscript}
\usepackage{enumerate}
\usepackage{slashed}
\usepackage{hyperref}
\usepackage{subfigure} 
\usepackage{epstopdf} 

\graphicspath{{./figures/}} 
\hypersetup{pdftex,colorlinks=true,linkcolor=blue,citecolor=blue,menucolor=black,urlcolor=blue,filecolor=blue}

\begin{document}

\title{Stairway to equilibrium entropy}

\author{Romulo Rougemont}
\email{rougemont@ufg.br}
\affiliation{Instituto de F\'{i}sica, Universidade Federal de Goi\'{a}s, Av. Esperan\c{c}a - Campus Samambaia, CEP 74690-900, Goi\^{a}nia, Goi\'{a}s, Brazil}

\author{Willians Barreto}
\email{willians.barreto@ufabc.edu.br}
\affiliation{Centro de Ci\^{e}ncias Naturais e Humanas, Universidade Federal do ABC, Av. dos Estados 5001, 09210-580 Santo Andr\'{e}, S\~{a}o Paulo, Brazil}
\affiliation{Centro de F\'{i}sica Fundamental, Universidad de Los Andes, M\'{e}rida 5101, Venezuela}

\begin{abstract}
We compute the time evolution of the non-equilibrium entropy in the homogeneous isotropization dynamics of the 1RCBH model, which has a critical point in its conformal phase diagram defined at finite temperature and R-charge density. We also evaluate the time evolution of the pressure anisotropy and the scalar condensate of the medium. We disclose a new feature (not present in the Bjorken flow dynamics analyzed in previous works), which is observed for all the analyzed initial data: the formation of a periodic sequence of several close plateaus in the form of a stairway for the entropy density near thermodynamic equilibrium. We find that the period of plateau formation in the stairway is half the period of oscillations of the slowest quasinormal mode of the system, which is therefore strongly tied to the late time dissipative dynamics of the system associated to the irreversibility of entropy production. For the particular case of the purely thermal SYM plasma at zero density and vanishing scalar condensate, we find that the period of the stairway is half the period of oscillations of the slowest quasinormal mode associated to the late time equilibration of the pressure anisotropy of the fluid, while at finite chemical potential the slowest quasinormal mode of the system is associated to the late time equilibration of the scalar condensate.
\end{abstract}

\maketitle
\tableofcontents


\section{Introduction}
\label{sec:intro}

The investigation of entropy production and thermalization in initially out of equilibrium dynamical systems is of fundamental interest for several areas of research in physics, see e.g. \cite{Rigol:2007juv,Rigol:2009zz,Muller:2011ra,Muller:2020ziz,Ebner:2023ixq,Gavassino:2021kjm,Herrera:2011kd}. In a broad sense, one could think on defining thermodynamic equilibrium for a dynamical system as a putative state in the late time evolution of the system for which entropy reaches its maximum value and becomes stationary. Indeed, the second law of thermodynamics states that for an isolated system the entropy either increases (for irreversible processes) or remains stationary (for reversible processes). If a given dynamical system thermalizes, at equilibrium there should be no net macroscopic flow of energy and matter within the system (at least for the time scales where the relevant observations are done). Therefore, due to the absence of gradients of temperature, energy and chemical composition between the different parts of the system, only reversible processes may take place and the entropy remains stationary at equilibrium. It is important to notice, however, that in some cases one may have transient time windows with zero entropy production even though the dynamical system under consideration is still far from equilibrium --- in such cases, although constant within a limited time interval, the entropy still increases at late times and may eventually approach a definitive stationary state where it reaches its maximum value for given values of internal energy and charge densities, see e.g. \cite{Rougemont:2021qyk,Rougemont:2021gjm,Rougemont:2022piu}. Furthermore, the equilibrium state, when it exists, should not depend on most details of the initial data evolved in time. Instead, it is characterized by a few macroscopic parameters like e.g. temperature ($T$) and chemical potential ($\mu$). For given values of these control parameters, a single stable equilibrium state should exist (besides possibly other unstable and metastable states), while several far from equilibrium initial data with different time evolutions may converge to a single stable equilibrium state characterized by a given value of $\mu/T$. Thus, in a certain sense, entropy production and thermalization effectively ``erase'' some details of the far from equilibrium initial states considered, leading to an effectively universal description of the long time, stationary behavior of the system in terms of just a few macroscopic control parameters.

Some subtleties may arise, however. In fact, as we are going to discuss in the present work, one may find that the entropy of some dynamical systems has already approximately equilibrated within some numerical tolerance while other physical observables (like the condensate of some scalar operator in a quantum field theory) still not reached its equilibrium value. Indeed, more generally, one may refer to the ``thermalization time'' of a dynamical system as the latest characteristic time scale for which all the relevant observables of the system have reached their equilibrium values within a given tolerance \cite{Critelli:2017euk}, and then one may refer to different characteristic equilibration times for different operators of a given quantum field theory \cite{Attems:2017zam}. This is interesting because one may then study how fast different physical observables equilibrate for different physical systems initially defined out of equilibrium. In particular, the holographic gauge-gravity duality \cite{Maldacena:1997re,Gubser:1998bc,Witten:1998qj,Witten:1998zw} allows for the calculation of several kinds of far from equilibrium dynamics in different strongly coupled quantum media, see e.g. \cite{Chesler:2008hg,Chesler:2009cy,Chesler:2010bi,Heller:2011ju,Heller:2012je,vanderSchee:2012qj,Casalderrey-Solana:2013aba,Chesler:2013lia,vanderSchee:2014qwa,Jankowski:2014lna,Fuini:2015hba,Chesler:2015bba,Bellantuono:2015hxa,Pedraza:2014moa,DiNunno:2017obv,Attems:2016tby,Casalderrey-Solana:2016xfq,Grozdanov:2016zjj,Attems:2017zam,Romatschke:2017vte,Spalinski:2017mel,Critelli:2017euk,Casalderrey-Solana:2017zyh,Critelli:2018osu,Attems:2018gou,Waeber:2019nqd,Kurkela:2019set,Cartwright:2019opv,Cartwright:2020qov,Rougemont:2021qyk,Rougemont:2021gjm,Cartwright:2021maz,Ecker:2021ukv,Ghosh:2021naw,Cartwright:2022hlg,Rougemont:2022piu} for a non-exhaustive list.

In Ref. \cite{Critelli:2017euk}, the homogeneous isotropization dynamics of the top-down holographic \textit{1 R-Charge Black Hole} (1RCBH) model \cite{Gubser:1998jb,Behrndt:1998jd,Kraus:1998hv,Cai:1998ji,Cvetic:1999ne,Cvetic:1999rb,DeWolfe:2011ts,DeWolfe:2012uv} was analyzed with the calculation of the time evolution of the pressure anisotropy and the scalar condensate for a set of initially homogeneous but anisotropic far from equilibrium states. In the present work, we numerically evaluate and analyze for the first time the evolution of the non-equilibrium entropy of the 1RCBH model undergoing homogeneous isotropization dynamics for several initial data, while also comparing with the results for the pressure anisotropy and the scalar condensate of the dynamical medium. Interestingly and akin to what has been previously found for the inhomogeneous Bjorken flow of the same model \cite{Rougemont:2022piu},\footnote{The 1RCBH model has as a particular case, with zero R-charge chemical potential and zero scalar condensate, the purely thermal $\mathcal{N}=4$ Supersymmetric Yang-Mills (SYM) plasma, whose Bjorken flow dynamics has also been shown to admit dynamical solutions with transient violations of energy conditions \cite{Rougemont:2021qyk,Rougemont:2021gjm}.} also for the homogeneous isotropization dynamics there are numerical solutions for which the corresponding initial data satisfy all the energy conditions, but whose dynamical evolutions transiently violate the dominant and even the weak energy conditions when the medium is far from equilibrium. On the other hand, contrary to what happens in the Bjorken flow, in the homogeneous isotropization dynamics of the 1RCBH model the violations of energy conditions are generally reduced as the ratio $\mu/T$ is increased.

As in the Bjorken flow, for some solutions we also observe the formation of transient plateaus in the non-equilibrium entropy, however, differently than in the Bjorken flow, in the homogeneous isotropization dynamics not all the transient plateaus are anticipating a posterior violation of energy conditions. In fact, we observe a more complex structure of transient plateaus in the homogeneous isotropization dynamics than in the Bjorken flow of the 1RCBH model.

Indeed, the main result of the present work regards the disclosure of a new feature corresponding to the formation of a periodic sequence of several close plateaus for the entropy density near thermodynamic equilibrium, which is observed for all the solutions analyzed. Such a structure resembles the form of a stairway as the entropy approaches its asymptotic equilibrium value. The near-equilibrium plateaus for the entropy may be so close to each other that they might go unnoticed at first glance, therefore they typically require a high numerical precision to be resolved. Interestingly, by adopting a not so small tolerance to define the characteristic equilibration times of different physical observables, one may effectively conclude that the entropy in the homogeneous isotropization dynamics of the 1RCBH model approximately equilibrates well before the scalar condensate, with the latter then setting the actual thermalization time for this system.\footnote{In fact, in Ref. \cite{Critelli:2017euk} it was argued that the equilibration of the scalar condensate should be regarded as the thermalization time of this system, since it was always observed to be much larger than the isotropization time associated with the (approximate) vanishing of the pressure anisotropy of the medium. The inclusion of the non-equilibrium entropy in the set of dynamical observables analyzed in the present work does not change this conclusion.} However, a subtlety remains in this regard, since the observation of the sequential close plateaus for the entropy density near equilibrium indicates that some intriguing and nontrivial physical effects are still taking place at such late stages in the evolution of the system. This becomes even more apparent by the fact that typically the scalar condensate still presents appreciable oscillations around its asymptotic equilibrium value even for considerably larger time scales. In fact, as one of the main results of the present work, we identify the period of plateau formation in the stairway to equilibrium entropy as half the period of oscillations of the slowest quasinormal mode of the system, which is associated to the equilibration of the scalar condensate for finite density states of the medium. For the particular case of the purely thermal SYM plasma at zero density and vanishing scalar condensate, we find that the period of the stairway is half the period of oscillations of the slowest quasinormal mode associated to the equilibration of the pressure anisotropy of the fluid. 

The present manuscript is structured as follows: in sections \ref{sec:2.1} and \ref{sec:2.2} we briefly review some of the main aspects used in the holographic computation of the homogeneous isotropization dynamics of the 1RCBH model \cite{Critelli:2017euk}. In section \ref{sec:2.3} we derive the holographic formula for the calculation of the holographic non-equilibrium entropy in this setup, while in section \ref{sec:2.4} we derive the energy conditions for the 1RCBH model undergoing homogeneous isotropization dynamics, besides also discussing the set of initial data analyzed in the present work. In section \ref{sec:3} we present our main results, with the outcomes for the time evolution of the pressure anisotropy, the scalar condensate, and the non-equilibrium entropy, with the latter revealing the stairway structure to equilibrium entropy which is present for all the numerical solutions analyzed. The conclusions and future perspectives are presented in section \ref{sec:conc}. 

In this work we use a mostly plus metric signature and natural units where $c=\hbar=k_B=1$.

\section{Homogeneous isotropization dynamics of the 1RCBH model}
\label{sec:2}

The 1RCBH model \cite{Gubser:1998jb,Behrndt:1998jd,Kraus:1998hv,Cai:1998ji,Cvetic:1999ne,Cvetic:1999rb,DeWolfe:2011ts,DeWolfe:2012uv} is a top-down gauge-gravity construction holographically dual to a strongly coupled and conformal $\mathcal{N}=4$ SYM plasma at finite temperature with a chemical potential associated to the conserved R-charge related to an Abelian $U(1)$ subgroup of the $SU(4)$ R-charge global symmetry of the boundary quantum field theory. Its bulk description is given in terms of a five dimensional Einstein-Maxwell-Dilaton (EMD) action,
\begin{align}
S = \frac{1}{2 \kappa_5^2} \int d^5x \sqrt{-g} \left[ R - \frac{f(\phi)}{4} F_{\mu \nu} F^{\mu \nu} - \frac{1}{2} (\partial_{\mu} \phi)^2 - V(\phi) \right],
\label{eq:action}
\end{align}
where $\kappa_5^2\equiv 8 \pi G_5$ is the five dimensional gravity constant, while the dilaton potential $V(\phi)$ and the Maxwell-dilaton coupling function $f(\phi)$ read,
\begin{align}
V(\phi) = -\frac{1}{L^2} \left(8 e^{\phi/\sqrt{6}} + 4 e^{-\sqrt{2/3}\,\phi} \right), \qquad f(\phi) = e^{- 2\sqrt{2/3}\,\phi},
\label{eq:Vandf}
\end{align}
where $L$ is the asymptotic AdS$_5$ radius (which we set here to unity). The bulk action \eqref{eq:action} is accompanied by two boundary actions: the traditional Gibbons-Hawking-York action \cite{York:1972sj,Gibbons:1976ue} required for the well-posedness of the Dirichlet boundary condition problem in spacetimes with boundaries \cite{Poisson:2009pwt} (as in the case of asymptotically AdS geometries), and a counterterm action \cite{Critelli:2017euk} constructed via the holographic renormalization procedure \cite{Bianchi:2001kw,Skenderis:2002wp,deHaro:2000vlm} with the purpose of consistently removing the boundary divergences of the full on-shell action.

Due to its simplicity and to the fact that it is a rigorous top-down holographic construction describing a strongly coupled quantum medium at finite temperature and density with a critical point in its phase diagram, the 1RCBH model is being widely explored in the holographic literature in recent years. Indeed, for instance, its thermodynamics has been analyzed in \cite{DeWolfe:2011ts,Finazzo:2016psx}, some hydrodynamic transport coefficients have been calculated in \cite{DeWolfe:2011ts,Asadi:2021hds}, the spectra of quasinormal modes have been obtained in \cite{Finazzo:2016psx,Critelli:2017euk}, several observables of quantum information theory were evaluated in \cite{Ebrahim:2018uky,Ebrahim:2020qif,Amrahi:2021lgh}, chaotic properties and the pole-skipping phenomenon have been addressed in \cite{Amrahi:2023xso,Pant:2023oky}, while the holographic renormalization and previous far from equilibrium numerical simulations of homogeneous isotropization dynamics and inhomogeneous Bjorken flow were discussed in \cite{Critelli:2017euk,Critelli:2018osu,Rougemont:2022piu}.

The focus of the present work is on the computation of the non-equilibrium entropy of the 1RCBH model undergoing homogeneous isotropization dynamics, besides the analysis of the energy conditions during the time evolution of the system. None of these topics have been addressed before in such a context and will be discussed in detail in the present work, but since most of the details of the homogeneous isotropization dynamics of the 1RCBH model have been already covered in Ref. \cite{Critelli:2017euk}, most of the discussion in the present section will be presented in the form of a brief revision for the convenience of the reader (for full details we refer the interested reader to consult \cite{Critelli:2017euk} and references therein).

\subsection{Late time equilibrium thermodynamics}
\label{sec:2.1}

In this section we briefly review the main points regarding the thermodynamics of the 1RCBH model required for the purposes of the present work. Thermodynamic equilibrium is only reached at late times in the evolution of the homogeneous isotropization dynamics starting from far from equilibrium anisotropic initial states.

The 1RCBH model is a conformal field theory with no intrinsic energy scale, therefore, its hot and dense phase diagram is one dimensional, being effectively described in terms of a single control parameter, the dimensionless ratio $\mu/T$, instead of independent $(T,\mu)$. The phase diagram of the 1RCBH model has the peculiar feature of being a limited line segment, $\mu/T\in[0,\pi/\sqrt{2}]$, where for $\mu/T=0$ (and zero scalar condensate) the theory reduces to the purely thermal SYM plasma, while at $\mu/T=\pi/\sqrt{2}$ there is a critical point where second (and higher) order derivatives of the pressure of the medium diverge \cite{DeWolfe:2011ts,Finazzo:2016psx}. However, since the phase diagram of the model ends at the critical point, there is no actual phase transition. For each value of $\mu/T$ there are two different solutions, one of which is thermodynamically unstable and another one which is stable, with the latter being of physical interest for us. In which branch of black hole solutions the system lies within is something controlled by the dimensionless ratio between the black hole charge and the radial position of its equilibrium event horizon, $Q/r_\textrm{EH}$ --- see Fig. 1 of \cite{Finazzo:2016psx}.

Within the thermodynamically stable branch of equilibrium black hole solutions, we shall use in this work two results that will serve as analytical consistency checks of the late time numerical evolution of the scalar condensate and the entropy density, besides serving also to characterize the effective equilibration times of these observables for different initial conditions. As in Ref. \cite{Critelli:2017euk}, we fix here the equilibrium temperature as $T=1/\pi$ and then measure the chemical potential of the medium with respect to this scale, setting $x\equiv\mu/T=\pi\mu$. By solving Eq. (4.10) of \cite{Critelli:2017euk} for the radial position of the event horizon in equilibrium and then substituting the result into Eq. (4.8) with the aforementioned choice for $T=1/\pi$, one finds the following result for the black hole charge in equilibrium as a function of $\mu/T$,
\begin{align}
Q(x\equiv\mu/T=\pi\mu) = \sqrt{4-\frac{(x/x_c)^2}{2}-\frac{2\, (x/x_c)^2}{1-\sqrt{1-(x/x_c)^2}}},
\label{eq:som3}
\end{align}
where $x_c=(\mu/T)_c=\pi/\sqrt{2}=\pi\mu_c$ gives the critical chemical potential. Letting $X$ be any physical observable and taking $\hat{X}\equiv\kappa_5^2 X = 4\pi^2 X/N_c^2$, by means of Eq. (4.24) of \cite{Critelli:2017euk} one finds that the thermodynamically stable equilibrium value of the normalized scalar condensate is given by,
\begin{align}
\frac{\langle \hat{O}_\phi\rangle_\textrm{eq}}{T^2} = \pi^2\,\sqrt{\frac{2}{3}}\,\,Q^2(x\equiv\mu/T=\pi\mu),
\label{eq:Ophieq}
\end{align}
with $Q(x\equiv\mu/T=\pi\mu)$ given by Eq. \eqref{eq:som3}. Moreover, by means of Eq. (4.11) of \cite{Critelli:2017euk} one obtains the following result for the thermodynamically stable equilibrium value of the normalized entropy density of the medium,
\begin{align}
\frac{\hat{s}_\textrm{eq}}{T^3} = \frac{\pi^4}{4} \left[3-\sqrt{1-\left(\frac{x}{x_c}\right)^2}\,\right]^2 \left[1+\sqrt{1-\left(\frac{x}{x_c}\right)^2}\,\right].
\label{eq:sT3eq}
\end{align}

\subsection{Holographic equations of motion out of equilibrium and renormalized 1-point functions}
\label{sec:2.2}

The general EMD equations of motion obtained by extremizing the bulk action \eqref{eq:action} are \cite{Critelli:2017euk,Rougemont:2022piu},
\begin{subequations}
\begin{align}
R_{\mu\nu}-\frac{g_{\mu\nu}}{3}\left(V(\phi)-\frac{f(\phi)}{4}F_{\alpha\beta}^2\right) -\frac{1}{2}\partial_\mu\phi\partial_\nu\phi-\frac{f(\phi)}{2}F_{\mu\rho}F_\nu^\rho &=0, \label{eq:Einstein}\\
\nabla_\mu(f(\phi)F^{\mu\nu})=\frac{1}{\sqrt{-g}}\partial_\mu(f(\phi)F^{\mu\nu}) &=0,
\label{eq:Maxwell}\\
\frac{1}{\sqrt{-g}}\partial_\mu(\sqrt{-g}g^{\mu\nu}\partial_\nu\phi) -\partial_\phi V(\phi) - \frac{\partial_\phi f(\phi)}{4}F_{\mu\nu}^2 &=0, \label{eq:dilaton}
\end{align}
\end{subequations}
while the ansatze for the bulk fields compatible with the symmetries of the homogeneous isotropization dynamics read as follows in generalized infalling Eddington-Finkelstein (EF) coordinates \cite{Critelli:2017euk},
\begin{equation}
ds^2= 2dv\left[dr-A(v,r) dv \right]+\Sigma(v,r)^2\left[e^{B(v,r)}(dx^2+dy^2)+e^{-2B(v,r)}dz^2\right],\,\,\,\, A_\mu dx^\mu = \Phi(v,r)dv,\,\,\,\, \phi = \phi(v,r),
\label{eq:ansatze}
\end{equation}
where $v$ is the EF time defined by the relation,
\begin{equation}
dv = dt +\sqrt{-\frac{g_{rr}}{g_{tt}}}dr.
\label{eq:EFtime}
\end{equation}
Since $g_{rr}$ and $g_{tt}$ are the holographic radial and temporal diagonal components of an asymptotically AdS$_{5}$ spacetime, as one approaches the boundary of the bulk geometry at $r\rightarrow\infty$, one obtains the time coordinate of the dual quantum field theory living at the boundary, $v\rightarrow t$. Infalling radial null geodesics satisfy $v=\textrm{constant}$ and outgoing radial null geodesics obey $dr/dv=A(v,r)$ \cite{Chesler:2008hg}.\footnote{Note that $A(v,r)$ here is half the corresponding metric function in the convention adopted in Ref. \cite{Chesler:2008hg}.} There is also a residual diffeomorphism invariance for the metric in \eqref{eq:ansatze} associated to the radial shift $r\mapsto r+\lambda(v)$, with $\lambda(v)$ being an arbitrary function of the EF time \cite{Chesler:2013lia}. Close to the boundary the bulk metric approaches AdS$_5$, the dilaton field approaches zero and the Maxwell field gives at asymptotically large times the R-charge chemical potential of the strongly coupled quantum fluid living at the boundary. The precise form of the boundary conditions for the bulk functions to be integrated will be specified in subsection \ref{sec:2.3}.

By substituting the particular ansatze \eqref{eq:ansatze} in the general EMD equations \eqref{eq:Einstein} --- \eqref{eq:dilaton}, one gets the following set of coupled $1+1$ partial differential equations \cite{Critelli:2017euk,Rougemont:2022piu},\footnote{We employed Eq. \eqref{eq:pde1} to write down the constraint \eqref{eq:Max-Const}, taking also into account that, from the definitions of $d_+$ and $\mathcal{E}$, one can rewrite $(d_+\Phi)' = -\partial_v\mathcal{E} - A'\mathcal{E} - A\mathcal{E}'$. In writing down the constraint \eqref{eq:EH-Const2} we employed the Hamiltonian constraint \eqref{eq:EH-HamiltConst}.}
\begin{subequations}
\begin{align}
\partial_v\mathcal{E}+A\mathcal{E}'+\left(3\frac{d_+\Sigma}{\Sigma}+\frac{\partial_\phi f}{f}d_+\phi\right)\mathcal{E} &=0,\label{eq:Max-Const}\\
\frac{3 \Sigma '}{\Sigma } + \frac{\partial_\phi f}{f}\phi ' + \frac{\mathcal{E}'}{\mathcal{E}} &=0,\label{eq:pde1}\\
4 \Sigma (d_+\phi)'+6 \phi'd_+\Sigma + 6 \Sigma'd_+\phi+\Sigma\mathcal{E}^2 \partial_\phi f - 2 \Sigma  \partial_\phi V &=0,\label{eq:pde2}\\
(d_+\Sigma)' +\frac{2 \Sigma '}{\Sigma }d_+\Sigma+\frac{\Sigma}{12} \left(2V + f \mathcal{E}^2\right) &=0,\label{eq:pde3}\\
\Sigma \, (d_{+}B)'+\frac{3}{2}(B' d_+\Sigma + \Sigma'd_+ B) &=0,\label{eq:pde4}\\
A'' + \frac{1}{12} \left(18 B'd_{+}B-\frac{72 \Sigma'd_{+}\Sigma}{\Sigma^2} + 6 \phi'd_{+}\phi-7 f \mathcal{E}^2-2 V\right) &=0,\label{eq:pde5}\\
\Sigma'' + \frac{\Sigma}{6} \left(3 \left(B'\right)^2+\left(\phi '\right)^2\right) &=0,\label{eq:EH-HamiltConst}\\
d_+(d_+\Sigma)+\frac{\Sigma}{2}(d_+B)^2-A'd_+\Sigma+\frac{\Sigma}{6}(d_+\phi)^2 &=0,\label{eq:EH-Const2}
\end{align}
\end{subequations}
where $X'\equiv \partial_r X$ is the directional derivative along infalling radial null geodesics, $d_+X\equiv[\partial_v+A(v,r)\partial_r]X$ is the directional derivative along outgoing radial null geodesics, and $\mathcal{E}\equiv-\Phi'$. Eqs. \eqref{eq:Max-Const} and \eqref{eq:pde1} are the nontrivial components of Maxwell's equation, Eq. \eqref{eq:pde2} is the dilaton equation, and Eqs. \eqref{eq:pde3} --- \eqref{eq:EH-Const2} are the nontrivial components of Einstein's equations. There are five unknown functions, $\{A(v,r),\Sigma(v,r),B(v,r),\phi(v,r),\mathcal{E}(v,r)\}$, to be determined by the five dynamical equations of motion \eqref{eq:pde1} --- \eqref{eq:pde5}, besides three constraints given by Eqs. \eqref{eq:Max-Const}, \eqref{eq:EH-HamiltConst}, and \eqref{eq:EH-Const2}. Eqs. \eqref{eq:pde1} --- \eqref{eq:EH-HamiltConst} form a nested set of equations of motion which may be numerically integrated, while the constraints \eqref{eq:Max-Const} and \eqref{eq:EH-Const2} may be employed to check the accuracy of such numerical solutions.

The form of the ultraviolet near-boundary expansions of the bulk fields for the 1RCBH model undergoing homogeneous isotropization dynamics may be written as follows \cite{Critelli:2017euk},

\begin{subequations}
\begin{align}
A(v,r) & = \frac{\left[r+\lambda(v)\right]^2}{2}-\partial_v\lambda(v) + \sum_{n=1}^{\infty}\frac{A_n(v)}{r^n}, \label{eq:expA}\\
\Sigma(v,r) & = r+\lambda(v) + \sum_{n=1}^{\infty}\frac{\Sigma_n(v)}{r^n}, \label{eq:expSig}\\
B(v,r) & = \sum_{n=1}^{\infty}\frac{B_n(v)}{r^n}, \label{eq:expB}\\
\phi(v,r) & = \sum_{n=2}^{\infty}\frac{\phi_n(v)}{r^n}, \label{eq:expphi}\\
\Phi(v,r) & = \Phi_0(v) + \sum_{n=2}^{\infty}\frac{\Phi_n(v)}{r^n}. \label{eq:expPhi}
\end{align}
\end{subequations}
As detailed discussed in \cite{Critelli:2017euk}, one may derive from the Maxwell's equation \eqref{eq:pde1} and from the ultraviolet near-boundary analysis of the bulk fields the following relation,
\begin{equation}
\mathcal{E}(v,r) = 2\Phi_{2}\Sigma(v,r)^{-3}e^{2\sqrt{\frac{2}{3}}\phi(v,r)}.
\label{eq:Efield2}
\end{equation}
Moreover, the physical observables related to the renormalized one-point functions of the boundary energy-momentum tensor and the charge current operator read as follows \cite{Critelli:2017euk},
\begin{subequations}
\begin{align}
\hat{\varepsilon} & \equiv \kappa_5^2\, \varepsilon = \kappa_5^2\, \langle T_{tt} \rangle = \kappa_5^2\, \varepsilon_\textrm{eq}(x=\mu/T) = \frac{3}{32} \left[3-\sqrt{1-\left(\frac{x}{x_c}\right)^2}\,\right]^3 \left[1+\sqrt{1-\left(\frac{x}{x_c}\right)^2}\,\right], \\
\hat{\rho} & \equiv \kappa_5^2\, \rho = \kappa_5^2\, \langle J^t \rangle = -\Phi_{2} = \kappa_5^2\, \rho_\textrm{eq}(x=\mu/T) = \frac{x}{4\pi} \left[3-\sqrt{1-\left(\frac{x}{x_c}\right)^2} \,\right]^2, \label{eq:rho}\\
\Delta \hat{p}(v) & \equiv \kappa_5^2\, (p_T-p_L) = \kappa_5^2\, \left[\langle T_{xx} \rangle - \langle T_{zz}\rangle\right] = 6B_{4}(v) , \label{eq:DeltaP}\\
\langle \hat{O}_{\phi} \rangle(v) & \equiv \kappa_5^2\, \langle O_\phi \rangle = -\phi_{2}(v), \label{eq:DilCondensate}
\end{align}
\end{subequations}
where $\kappa_5^2=4\pi^2/N_c^2$ for a strongly coupled SYM plasma (as the 1RCBH model), $\varepsilon$ is the internal energy density and $\rho$ is the R-charge density, both of which are conserved in the homogeneous isotropization dynamics, and $\Delta p$ is the pressure anisotropy of the medium.

As discussed in \cite{Critelli:2017euk}, one may fix the vast majority of the ultraviolet expansion coefficients in Eqs. \eqref{eq:expA} --- \eqref{eq:expPhi} in terms of just a few undetermined coefficients and their time derivatives. This is accomplished by substituting
these expansions in the EMD equations of motion and then solving the obtained algebraic equations order by order in powers of $r$. By considering ultraviolet expansions up to order $n=8$ there remain five undetermined coefficients in such analysis: $\{\Phi_0(v),\Phi_2(v),A_2(v),B_4(v),\phi_2(v)\}$. The coefficient $\Phi_{0}(v)$ is fixed by the Dirichlet boundary condition for the (nonzero time component of the) Maxwell field,
\begin{align}
\lim_{v\rightarrow\infty} \lim_{r\to\infty} \Phi(v,r)=\lim_{v\rightarrow\infty}\Phi_{0}(v)=\mu,
\end{align}
which gives the $U(1)$ R-charge chemical potential at the boundary quantum field theory. The coefficient $\Phi_2(v)$ is actually constant and it is related to the conserved R-charge density of the fluid as given in Eq. \eqref{eq:rho}. The coefficient $A_2(v)$ may be fixed in terms of the constant $H\equiv A_2(v)+\phi_2^2(v)/18=-\hat{\varepsilon}/3$ if one knows the coefficient $\phi_2(v)$. In fact, the two remaining undetermined ultraviolet coefficients $\{B_4(v),\phi_2(v)\}$ are dynamical quantities related to the scalar condensate and the pressure anisotropy according to Eqs. \eqref{eq:DeltaP} and \eqref{eq:DilCondensate}. The values of these two coefficients at the initial time slice can be freely chosen since they are given by the boundary values of the initial profiles for the metric anisotropy function $B(v,r)$ and the dilaton field $\phi(v,r)$, which are two of the three initial data that must be chosen for the 1RCBH model undergoing isotropization dynamics (the third initial data is the value of $\mu/T$), as we shall discuss in a moment; and once their initial values are specified, their subsequent time evolutions are determined by numerically solving the nested set of partial differential equations previously obtained.

Schematically, the numerical integration of the nested set of $1+1$ partial differential equations of motion describing the homogeneous isotropization dynamics of the 1RCBH model proceeds as follows:
\begin{enumerate}[a)]
\item On the hypersurface at the initial time slice $v_0$ (which we set to be zero here, $v_0=0$), choose the initial profiles for the metric anisotropy function $B(v_0,r)$ and for the dilaton field $\phi(v_0,r)$, besides also the value for the dimensionless ratio $\mu/T$ defining the chemical potential of the medium at the boundary, which is associated to the charge of the black hole solution within the bulk;\footnote{If one chooses to work with nonzero radial shift function $\lambda(v)$, also its initial value must be chosen, and we set here $\lambda(v_0=0)=0$; its time evolution can be obtained by requiring that the radial position of the apparent horizon of the black hole solution remains fixed during the time evolution of the system \cite{Chesler:2013lia}.}
\item Next radially solve the Hamiltonian constraint \eqref{eq:EH-HamiltConst} to obtain $\Sigma(v_0,r)$, which at this step fixes the value of $\mathcal{E}(v_0,r)$ through Eq. \eqref{eq:Efield2};
\item Next radially solve Eq. \eqref{eq:pde3} to obtain $d_+\Sigma(v_0,r)$;
\item Next radially solve Eq. \eqref{eq:pde4} to obtain $d_+B(v_0,r)$;
\item Next radially solve the dilaton Eq. \eqref{eq:pde2} to obtain $d_+\phi(v_0,r)$;
\item Next radially solve Eq. \eqref{eq:pde5} to obtain $A(v_0,r)$;
\item At this step, from the definition of the directional derivative along outgoing radial null geodesics, $d_+\equiv\partial_v+A(v,r)\partial_r$, one has $\{\partial_v B(v_0,r),\partial_v \phi(v_0,r)\}$, which together with the initial profiles chosen for the metric anisotropy and the dilaton field, comprise the set of initial conditions required to evolve $\{B(v_0,r),\phi(v_0,r)\}$ to the next time slice $v_0+\Delta v$ using discrete numerical integration techniques (here we employ the pseudospectral method \cite{boyd01} to perform the radial integrations, while the time integrations are performed with the 4th order Adams-Bashforth method);
\item Repeat the previous steps to obtain all the bulk functions in the current time slice and iterate the procedure until reaching any desired end time $v_\textrm{end}$ for the numerical simulations (for the calculations performed in the present work, we set $v_\textrm{end}=13$, which gives the dimensionless time measure $v_\textrm{end}T=13/\pi$).
\end{enumerate}

\subsection{Subtracted fields, boundary conditions, apparent horizon and the holographic non-equilibrium entropy}
\label{sec:2.3}

In Ref. \cite{Critelli:2017euk} it was considered the formulation of the homogeneous isotropization dynamics of the 1RCBH model with $\lambda(v)=0$. The physics does not depend on the choice of $\lambda(v)$ (since it works like a gauge function), however, depending on the system under consideration, for numerical stability the introduction of this function in the formalism may be required \cite{Chesler:2013lia}. This was the case for the Bjorken flow dynamics of the 1RCBH model \cite{Rougemont:2022piu}. For the homogeneous isotropization dynamics one does not really need to work with a nonzero radial shift function, nonetheless, for completeness, in the present work we consider a nontrivial $\lambda(v)$. We explicitly checked that the results for all the physical observables at the boundary quantum field theory are the same obtained with vanishing $\lambda(v)$, as it should be.

By considering the ultraviolet near-boundary expansions of the bulk fields with nonzero $\lambda(v)$ and by introducing a new compact radial coordinate $u\equiv 1/r$ suited for numerical integration, we introduce below subtracted bulk fields with the purpose of obtaining radial constants as the boundary values of the subtracted fields to be numerically integrated. We define $u^p X_s(v,u)\equiv X(v,u)-X_\textrm{UV}(v,u)$, where $p\in\mathbb{Z}$ and $X_\textrm{UV}(v,u)$ is some ultraviolet truncation of the field $X(v,u)$ such that $X_s(v,u=0)$ gives a radial constant. The subtracted fields to be numerically integrated are defined here as follows (note that in the equations below we also provide the boundary conditions for the subtracted fields),\footnote{One substitutes the expansions \eqref{eq:expA} --- \eqref{eq:expPhi} into the equations of motion \eqref{eq:pde1} --- \eqref{eq:EH-HamiltConst}, eliminates all possible coefficients in favor of the others, and then passes from the original radial coordinate $r$ to the new compact radial coordinate $u=1/r$.}
\begin{subequations}
\begin{align}
u^2 A_s(v,u) &\equiv A(v,u) - \frac{1}{2u^2} - \frac{\lambda(v)}{u} - \frac{\left[\lambda^2(v)-2\partial_v\lambda(v)\right]}{2}, \label{eq:14a}\\
\Rightarrow A_s(v,u\to 0) &\to \frac{18 H - \phi_2^2(v)}{18} - \frac{\left[\phi_2(v)\partial_v\phi_2(v)+\lambda(v)\left(36 H-2\phi_2^2(v)\right)\right] u}{18} +\mathcal{O}\left(u^2\right), \label{eq:14b}\\
\Rightarrow \partial_u A_s(v,u=0) &= -\frac{\phi_2(v) \left[\partial_u\phi_s(v,u=0)+2\lambda(v)\phi_2(v)\right]+\lambda(v)\left(36 H-2\phi_2^2(v)\right)}{18}, \label{eq:14c}
\end{align}
\end{subequations}
where we made use of Eq. \eqref{eq:17c} in obtaining Eq. \eqref{eq:14c} (which is used as an extra boundary condition in the radial integration of $A_s$) from Eq. \eqref{eq:14b},

\begin{subequations}
\begin{align}
u^4 B_s(v,u) &\equiv B(v,u), \label{eq:15a}\\
\Rightarrow B_s(v,u\to 0) &\to B_4(v) + \left[\partial_v B_4(v)-4\lambda(v) B_4(v)\right] u +\mathcal{O}\left(u^2\right), \label{eq:15b}\\
\Rightarrow \partial_v B_4(v) &= \partial_u B_s(v,u=0) + 4\lambda(v) B_4(v), \label{eq:15c}
\end{align}
\end{subequations}

\begin{subequations}
\begin{align}
u^2 \Sigma_s(v,u) &\equiv \Sigma(v,u) - \frac{1}{u} - \lambda(v), \label{eq:16a}\\
\Rightarrow \Sigma_s(v,u\to 0) &\to -\frac{\phi_2^2(v) u}{18} - \frac{\left[3\phi_2(v)\partial_v\phi_2(v)-5\lambda(v)\phi_2^2(v)\right] u^2}{30} +\mathcal{O}\left(u^3\right), \label{eq:16b}
\end{align}
\end{subequations}

\begin{subequations}
\begin{align}
u^2 \phi_s(v,u) &\equiv \phi(v,u), \label{eq:17a}\\
\Rightarrow \phi_s(v,u\to 0) &\to \phi_2(v) + \left[\partial_v \phi_2(v)-2\lambda(v) \phi_2(v)\right] u +\mathcal{O}\left(u^2\right), \label{eq:17b}\\
\Rightarrow \partial_v \phi_2(v) &= \partial_u \phi_s(v,u=0) + 2\lambda(v) \phi_2(v), \label{eq:17c}
\end{align}
\end{subequations}

\begin{align}
\mathcal{E}_s(v,u) \equiv \mathcal{E}(v,u), \label{eq:18}
\end{align}

\begin{subequations}
\begin{align}
u^2 \left(d_+\Sigma\right)_s(v,u) &\equiv \left(d_+\Sigma\right)(v,u) - \frac{1}{2u^2} - \frac{\lambda(v)}{u}-\frac{\lambda^2(v)}{2}, \label{eq:19a}\\
\Rightarrow \left(d_+\Sigma\right)_s(v,u\to 0) &\to H + \frac{\phi_2^2(v)}{36} +\mathcal{O}\left(u\right), \label{eq:19b}
\end{align}
\end{subequations}

\begin{subequations}
\begin{align}
u^3 \left(d_+B\right)_s(v,u) &\equiv \left(d_+B\right)(v,u), \label{eq:20a}\\
\Rightarrow \left(d_+B\right)_s(v,u\to 0) &\to -2B_4(v) +\mathcal{O}\left(u\right), \label{eq:20b}
\end{align}
\end{subequations}

\begin{subequations}
\begin{align}
u \left(d_+\phi\right)_s(v,u) &\equiv \left(d_+\phi\right)(v,u), \label{eq:21a}\\
\Rightarrow \left(d_+\phi\right)_s(v,u\to 0) &\to -\phi_2(v) +\mathcal{O}\left(u^2\right). \label{eq:21b}
\end{align}
\end{subequations}

The equations of motion to be numerically solved as functions of the coordinates $(v,u)$ are obtained by rewriting the original equations of motion in terms of the subtracted fields, whose boundary values were derived above. As detailed discussed in section 5.4 of \cite{Critelli:2017euk}, by discretizing the radial part of these continuum differential equations of motion using the pseudospectral method, one obtains an eigenvalue problem where one needs to invert a diagonal $(N-1)\times (N-1)$ matrix for each of the bulk fields, with $N$ being the number of collocation points of the radial Chebyshev-Gauss-Lobatto grid. These matrices correspond to the homogeneous part of the discretized differential equations of motion evaluated at each radial grid point, with the exception of the boundary grid point. The multiplication of the inverse matrices by the column vectors corresponding to the inhomogeneous part of the equations of motion provides the numerical solutions for the bulk fields. At the boundary grid point one must impose the boundary conditions derived above for the subtracted bulk fields. Then one must join to the $(N-1)$-dimensional eigenvectors obtained as solutions of the aforementioned eigenvalue problem the values of the respective bulk fields determined at the boundary grid point by the associated boundary conditions. In this way one obtains the complete $N$-dimensional eigenvectors providing the numerical solutions for the radial part of the EMD equations of motion at a hypersurface defined at any given time slice (the components of these $N$-dimensional eigenvectors are the values of the bulk fields at each of the $N$ collocation points of the radial grid).

The set of initial data needed to be chosen in order to start the time evolution of the system of partial differential equations of motion is $\{B_s(v_0,u),\phi_s(v_0,u),\mu/T\}$, besides also the value of $\lambda(v_0)$ if one wants to use a nontrivial radial shift function $\lambda(v)$. As aforementioned, we set here $\lambda(v_0=0)=0$. The equation of motion for $\lambda(v)$ comes from Eq. \eqref{eq:14a} evaluated at the radial location of the apparent horizon, $u = u_\textrm{AH} = 1/r_\textrm{AH}$, which for any metric of the form shown in Eq. \eqref{eq:ansatze} is obtained as the outermost solution of the equation $\left(d_+\Sigma\right)(v,r_\textrm{AH})=0$ \cite{Chesler:2013lia}, where one imposes that the radial location of the apparent horizon stays fixed during the time evolution of the system by requiring that $\partial_v r_\textrm{AH}=0\,\Rightarrow\, d_+\left[d_+\Sigma\right](v,r_\textrm{AH}) = A(v,r_\textrm{AH}) \,\partial_r \left[d_+\Sigma\right](v,r_\textrm{AH})$, which leads to the following condition when substituted in the constraint \eqref{eq:EH-Const2},
\begin{align}
A(v,u_{\textrm{AH}}) = \frac{6([d_+B](v,u_{\textrm{AH}}))^2 + 2([d_+\phi](v,u_{\textrm{AH}}))^2}{2V(\phi)+f(\phi)\mathcal{E}^2}.
\label{eq:Astar}
\end{align}
Substituting \eqref{eq:Astar} in Eq. \eqref{eq:14a} one obtains the equation of motion for the radial shift function,
\begin{align}
\partial_v\lambda(v) = u_{\textrm{AH}}^2 A_s(v,u_{\textrm{AH}})+\frac{1}{2u_{\textrm{AH}}^2}+\frac{\lambda(v)}{u_{\textrm{AH}}}+\frac{\lambda^2(v)}{2} -A(v,u_{\textrm{AH}}).
\label{eq:dlambda}
\end{align}
Eq. \eqref{eq:dlambda} evolves in time the initial condition $\lambda(v_0)$ by shifting the radial coordinate $u$ at each time slice such as to keep $u_\textrm{AH} = \textrm{constant}$. The apparent horizon is the outermost trapped null surface inside the event horizon and it (usually) converges to the latter at late times, when the black hole geometry approaches thermodynamic equilibrium. Since the apparent horizon is inside the event horizon, by cutting off the radial integration of the bulk equations of motion at some position inside the apparent horizon, one guarantees that the radial domain of the bulk spacetime causally connected to observers at the boundary is being properly taken into account and, consequently, no physical information is lost in this integration procedure.

In order to evolve in time the initial data $\{B_s(v_0,u),\phi_s(v_0,u)\}$ one needs to obtain the time derivatives $\partial_v B_s$ and $\partial_v \phi_s$, what can be done by taking the expressions for $d_+B=\partial_v B+A\partial_r B$ and $d_+\phi=\partial_v \phi+A\partial_r \phi$ rewritten in terms of the subtracted bulk fields and the compact radial coordinate $u=1/r$. One then obtains the following results,
\begin{subequations}
\begin{align}
\partial_v B_s(v,u) &= \frac{[d_+B]_s}{u} + \frac{2B_s}{u} + \frac{B_s'}{2} + 4u^3A_sB_s + u^4A_sB_s' + \left(4B_s+uB_s'\right) \lambda + \left(2uB_s +\frac{u^2B_s'}{2} \right)\lambda^2 \nonumber\\
& \,\,\,\,\,\,\, - \left( 4uB_s + u^2B_s' \right)\partial_v\lambda, \label{eq:dtBs}\\
\partial_v B_s(v,u=0) &= \partial_v B_4(v), \label{eq:dtBs0}
\end{align}
\end{subequations}

\begin{subequations}
\begin{align}
\partial_v\phi_s(v,u) & = \frac{[d_+\phi]_s}{u} + \frac{\phi_s}{u} + \frac{\phi_s'}{2} +2u^3A_s\phi_s +u^4A_s\phi_s' + \left(u\phi_s'+2\phi_s\right)\lambda +\left(u\phi_s+\frac{u^2\phi_s'}{2}\right)\lambda^2 \nonumber\\
& \,\,\,\,\,\,\, -\left(2u\phi_s+u^2\phi_s'\right)\partial_v\lambda, \label{eq:dtphis}\\
\partial_v \phi_s(v,u=0) &= \partial_v \phi_2(v), \label{eq:dtphis0}
\end{align}
\end{subequations}
with $X_s'(v,u)\equiv \partial_u X_s(v,u)$ being calculated at any fixed time slice by applying the pseudospectral finite differentiation matrix \cite{boyd01} to the numerical solution $X_s(v,u)$.

Now we derive the holographic formula for the non-equilibrium entropy density of the strongly coupled quantum fluid living at the boundary of the bulk geometry. The famous Bekenstein-Hawking's equation \cite{Bekenstein:1973ur,Hawking:1974sw} relates the area of the event horizon of a black hole in equilibrium with its entropy and, through the holographic dictionary, this gives the thermodynamic entropy of the dual quantum field theory also in equilibrium. However, for strongly coupled media out of equilibrium, it has been argued e.g. in Ref. \cite{Figueras:2009iu} that the corresponding holographic non-equilibrium entropy should be calculated from the area of the apparent horizon instead of the area of the event horizon. This is so because one expects that entropy production is local in time, therefore, associating the out of equilibrium entropy to the area of a dynamic event horizon seems rather unnatural because in such a context the event horizon can only be determined with the knowledge of the entire future evolution of the black hole geometry, consequently, it is a global rather than a local observable.\footnote{Ref.\ \cite{Figueras:2009iu} also provides a strong counterexample to further justify such a proposal: for the conformal soliton flow \cite{Friess:2006kw}, corresponding to an ideal fluid, entropy production must be identically zero at all times. The entropy calculated from the area of the apparent horizon in such a case is indeed constant, however, the area of the event horizon diverges showing that it is an inadequate measure of the non-equilibrium entropy of the dual medium at least in this case. In fact, for the conformal soliton flow the system does not approach a stationary state at late times, and the apparent horizon does not converge to the event horizon, differently from what happens in systems where dissipation drives the evolution of the medium, as in the 1RCBH model analyzed in the present work.} Also in several other works in the literature \cite{Chesler:2009cy,Heller:2011ju,Heller:2012je,Jankowski:2014lna,vanderSchee:2014qwa,Grozdanov:2016zjj,Buchel:2016cbj,Muller:2020ziz,Engelhardt:2017aux} the holographic non-equilibrium entropy has been associated to the area of the apparent horizon, and here we adopt the same approach. Note that since the apparent horizon is inside the event horizon and, for the 1RCBH model it does converge to the event horizon at late times, for sufficiently long times the areas of both horizons coincide and provide the same result for the entropy of the dual medium in thermodynamic equilibrium.

The area of the black hole apparent horizon in infalling EF coordinates is given by,
\begin{align}
A_{\textrm{AH}}(v) = \int_\textrm{AH} d^3x\, \sqrt{|\gamma_\textrm{AH}|}\,\biggr|_{u=u_{\textrm{AH}}} = \int_\textrm{AH} dx dy dz \,\sqrt{g_{xx}g_{yy}g_{zz}}\,\biggr|_{u=u_{\textrm{AH}}} = |\Sigma(v,u_{\textrm{AH}})|^3 \,V_3,
\label{eq:AreaHor}
\end{align}
where $V_3 = \int_\textrm{AH} dx dy dz$ is the volume of the planar horizon parallel to the boundary. Resorting to an analogous expression to the Bekenstein-Hawking's relation, one sets for the holographic non-equilibrium entropy of the fluid,
\begin{align}
\hat{s}_{\textrm{AH}}(v) \equiv \kappa_5^2\, s_{\textrm{AH}}(v) = \kappa_5^2\, \frac{S_{\textrm{AH}}(v)}{V_3} = \kappa_5^2\, \frac{A_{\textrm{AH}}(v)/4G_5}{V_3} = 2\pi|\Sigma(v,u_{\textrm{AH}})|^3,
\label{eq:hats}
\end{align}
where, from Eq. \eqref{eq:16a},
\begin{align}
\Sigma(v,u_\textrm{AH}) = u^2_\textrm{AH} \Sigma_s(v,u_\textrm{AH}) + \frac{1}{u_\textrm{AH}} + \lambda(v).
\label{eq:sigmauAH}
\end{align}
Therefore, setting $T=1/\pi$ as before, one has for the normalized non-equilibrium entropy density of the medium,
\begin{align}
\frac{\hat{s}_{\textrm{AH}}(v)}{T^3} = 2\pi^4 |\Sigma(v,u_{\textrm{AH}})|^3,
\label{eq:sT3}
\end{align}
which must converge to the analytical result \eqref{eq:sT3eq} at late times for any chosen value of $\mu/T$. This in fact happens for all the initial data we analyzed, as will be shown in section \ref{sec:3}.

\subsection{Energy conditions and initial data}
\label{sec:2.4}

Now, by following a similar approach as the one devised in \cite{Janik:2005zt}, we derive the weak energy condition (WEC) and the dominant energy condition (DEC) for the conformal 1RCBH model undergoing homogeneous isotropization dynamics. These classical energy conditions are usually postulated in general relativity to restrict the content of the energy-momentum tensor of matter employed in Einstein's equations with the aim of enforcing energy positiveness, although some quantum effects are known to violate these energy conditions \cite{Visser:1999de,Costa:2021hpu}.

The WEC states that $\langle \hat{T}_{\mu\nu}\rangle t^\mu t^\nu\ge 0$ for any timelike vector $t^\mu$,\footnote{We note that for a conformal fluid, like the 1RCBH model, the strong energy condition (SEC), which states that $\langle\hat{T}_{\mu\nu}\rangle t^\mu t^\nu\ge -\langle \hat{T}_\mu^\mu\rangle /2$, is equivalent to the WEC, since $\langle\hat{T}_\mu^\mu\rangle=0$ for a conformal medium.} while the DEC posits that for any future-directed timelike vector $t^\mu$ (i.e. with $t^v>0$), $X^\mu \equiv -\langle\hat{T}^{\mu\nu}\rangle t_\nu$ must also be a future-directed timelike or null vector (this is a sufficient but not a necessary condition to establish causal propagation of matter \cite{Wald:1984rg}).

For the homogeneous isotropization dynamics of the 1RCBH model, one has from Eqs. (4.40) --- (4.42) and (4.45) of \cite{Critelli:2017euk} the following form for the boundary energy-momentum tensor,
\begin{align}
\langle\hat{T}_{\mu\nu}\rangle = \textrm{diag}\left(\hat{\varepsilon},\hat{p}_T,\hat{p}_T,\hat{p}_L\right) = \textrm{diag}\left(\hat{\varepsilon},\frac{\hat{\varepsilon}+\Delta\hat{p}}{3},\frac{\hat{\varepsilon}+\Delta\hat{p}}{3},\frac{\hat{\varepsilon}-2\Delta\hat{p}}{3}\right).
\label{eq:Tmunu}
\end{align}
On the other hand, the most general timelike or null vector at the flat boundary may be written as, $t^\mu\equiv\left(\sqrt{s^2+2\omega^2+\xi^2},\omega,\omega,\xi\right)$. In fact, this implies that, $t_\mu t^\mu = -(s^2+2\omega^2+\xi^2)+2\omega^2+\xi^2=-s^2\le 0,\,\forall\,s,\omega,\xi\in\mathbb{R}$.

Now we analyze the WEC,
\begin{align}
0 \le \langle\hat{T}_{\mu\nu}\rangle t^\mu t^\nu = \hat{\varepsilon} s^2 + \left(4\hat{\varepsilon}+\Delta\hat{p}\right)\frac{2\omega^2}{3} + \left(2\hat{\varepsilon}-\Delta\hat{p}\right)\frac{2\xi^2}{3},
\label{eq:wec1}
\end{align}
and since $\{s,\omega,\xi\}$ are arbitrary real numbers, \eqref{eq:wec1} can only be satisfied $\forall\,s,\omega,\xi\in\mathbb{R}$ if,
\begin{equation}
\boxed{\hat{\varepsilon}\ge 0}\,;\qquad
\left.
    \begin{array}{lr}
        4\hat{\varepsilon}+\Delta\hat{p} &\ge 0 \\
        2\hat{\varepsilon}-\Delta\hat{p} &\ge 0
    \end{array}
\right\} \Rightarrow \boxed{-4\le \frac{\Delta\hat{p}}{\hat{\varepsilon}}\le 2}\,.
\label{eq:wec2}
\end{equation}

For the DEC one has that,
\begin{align}
0 &\le X^v = -\langle \hat{T}^{vv} \rangle t_v = \hat{\varepsilon} \sqrt{s^2+2\omega^2+\xi^2} \Rightarrow \boxed{\hat{\varepsilon} > 0}\,, \label{eq:dec1}\\
0 &\ge X_\mu X^\mu = -s^2 \left[\hat{\varepsilon}^2\right] -2\omega^2 \left[\hat{\varepsilon}^2-\left(\frac{\hat{\varepsilon}+\Delta\hat{p}}{3}\right)^2\right] - \xi^2 \left[\hat{\varepsilon}^2-\left(\frac{\hat{\varepsilon}-2\Delta\hat{p}}{3}\right)^2\right], \label{eq:dec2}
\end{align}
and since \eqref{eq:dec2} must be valid $\forall\,s,\omega,\xi\in\mathbb{R}$, it follows that,
\begin{align}
0 &\le 1 - \left(\frac{1+\Delta\hat{p}/\hat{\varepsilon}}{3}\right)^2 \Rightarrow \left|1+\frac{\Delta\hat{p}}{\hat{\varepsilon}}\right| \le 3 \Rightarrow -4\le \frac{\Delta\hat{p}}{\hat{\varepsilon}}\le 2, \label{eq:dec3}\\
0 &\le 1 - \left(\frac{1-2\Delta\hat{p}/\hat{\varepsilon}}{3}\right)^2 \Rightarrow \left|1-\frac{2\Delta\hat{p}}{\hat{\varepsilon}}\right| \le 3 \Rightarrow \boxed{-1\le \frac{\Delta\hat{p}}{\hat{\varepsilon}}\le 2}\,, \label{eq:dec4}
\end{align}
where \eqref{eq:dec4} is clearly more restrictive than \eqref{eq:dec3}.

Interestingly, for the homogeneous isotropization dynamics of the 1RCBH model, the WEC, given by the conditions in \eqref{eq:wec2}, and the DEC, given by the conditions in \eqref{eq:dec1} and \eqref{eq:dec4}, have the same form obtained in the case of the Bjorken flow dynamics \cite{Rougemont:2021qyk,Rougemont:2021gjm,Rougemont:2022piu}.

We remark that although the DEC and the WEC may be transiently violated in strongly coupled far-from-equilibrium quantum systems depending on the initial data being evolved in time, the quantum null energy condition (QNEC) \cite{Bousso:2015mna} is expected to be satisfied in general, including the cases of holographic systems defined far-from-equilibrium \cite{Ecker:2017jdw}.\footnote{In the present work we compute the out of equilibrium entropy associated to the area of the apparent horizon, as aforementioned, but still do not analyze the QNEC, which requires calculating functional derivatives of the entanglement entropy of the system, which is not being evaluated here.}

Finally, the set of initial conditions analyzed in the present work is given by the following functional forms for the initial profiles of the metric anisotropy function and the dilaton field,
\begin{align}
B_s(v_0=0,u) &= \mathcal{B}\, e^{-\mathcal{S}_B (u-u_B)^2}, \label{eq:Bs0}\\
\phi_s(v_0=0,u) &= \mathcal{F}\, e^{-\mathcal{S}_\phi (u-u_\phi)^2}, \label{eq:phis0}
\end{align}
where the parameters chosen in the present work are given in Table \ref{tabICs}.

\begin{table}[h]
\centering
\begin{tabular}{|c||c|c|c||c|c|c|}
\hline
IC $\#$ & $\mathcal{B}$ & $\mathcal{S}_B$ & $u_B$ & $\mathcal{F}$ & $\mathcal{S}_\phi$ & $u_\phi$ \\
\hline
\hline
1 & 0.1 & $10^2$ & 0.4 & $-0.01$ & $10^2$ & 0.3 \\
\hline
2 & 0.1 & 0 & 0 & $-0.1$ & 1 & 0.3 \\
\hline
3 & 0.5 & $10^2$ & 0.4 & $-\sqrt{2/3}\, Q^2$ & 0 & 0 \\
\hline
4 & 0.5 & 10 & 0.4 & $-\sqrt{2/3}\, Q^2$ & 0 & 0 \\
\hline
\end{tabular}
\caption{Set of parameters for the initial profiles of the subtracted metric anisotropy function \eqref{eq:Bs0} and the subtracted dilaton field \eqref{eq:phis0}.}
\label{tabICs}
\end{table}

\noindent For each pair of initial profiles $\{B_s,\phi_s\}$ specified above we consider some values for the ratio between the chemical potential and the temperature, $\mu/T$ , as discussed in the next section.

\section{Time evolution of the system and the stairway to equilibrium entropy}
\label{sec:3}

\begin{figure*}
\center
\subfigure[]{\includegraphics[width=0.49\textwidth]{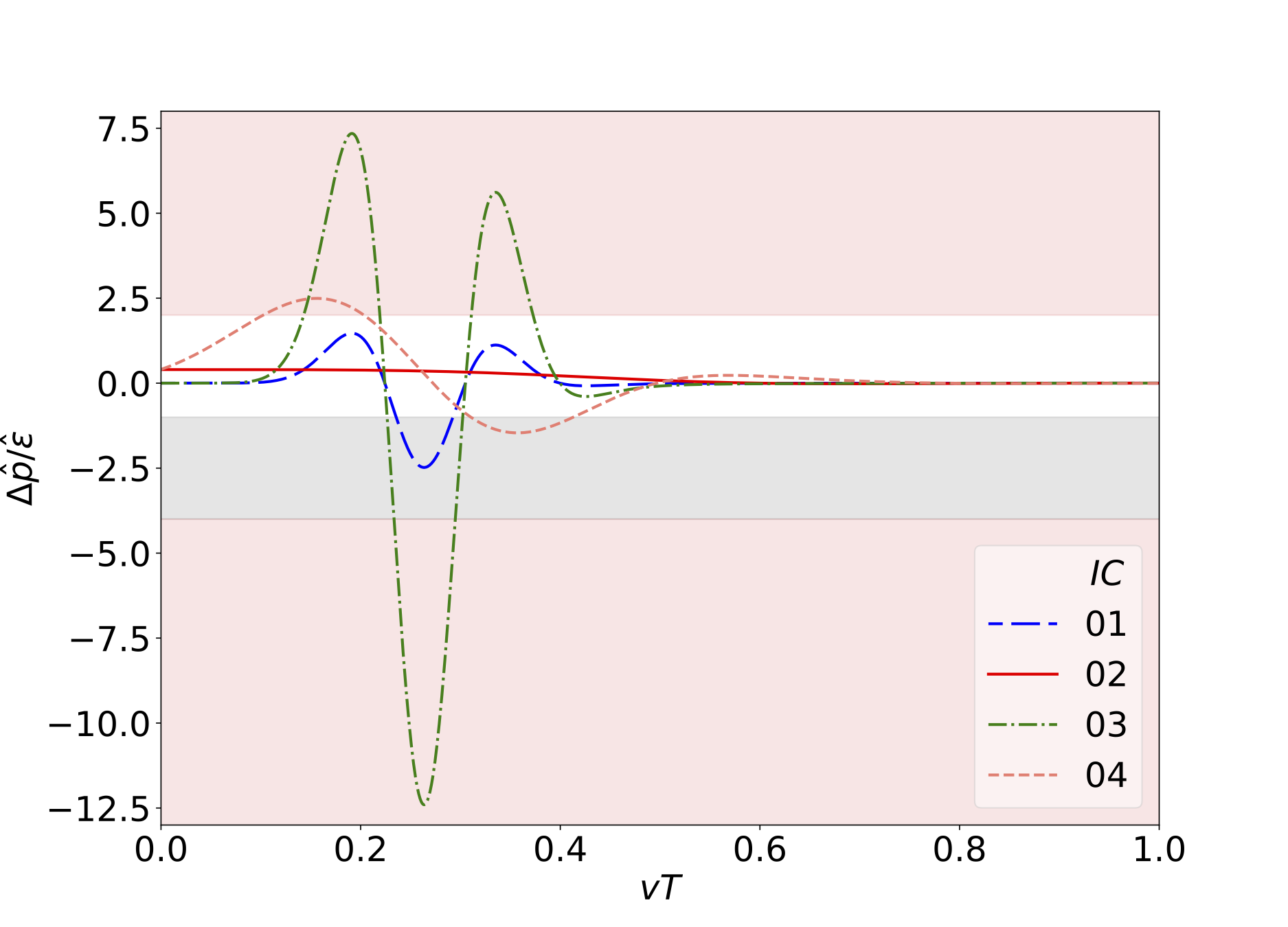}}
\subfigure[]{\includegraphics[width=0.49\textwidth]{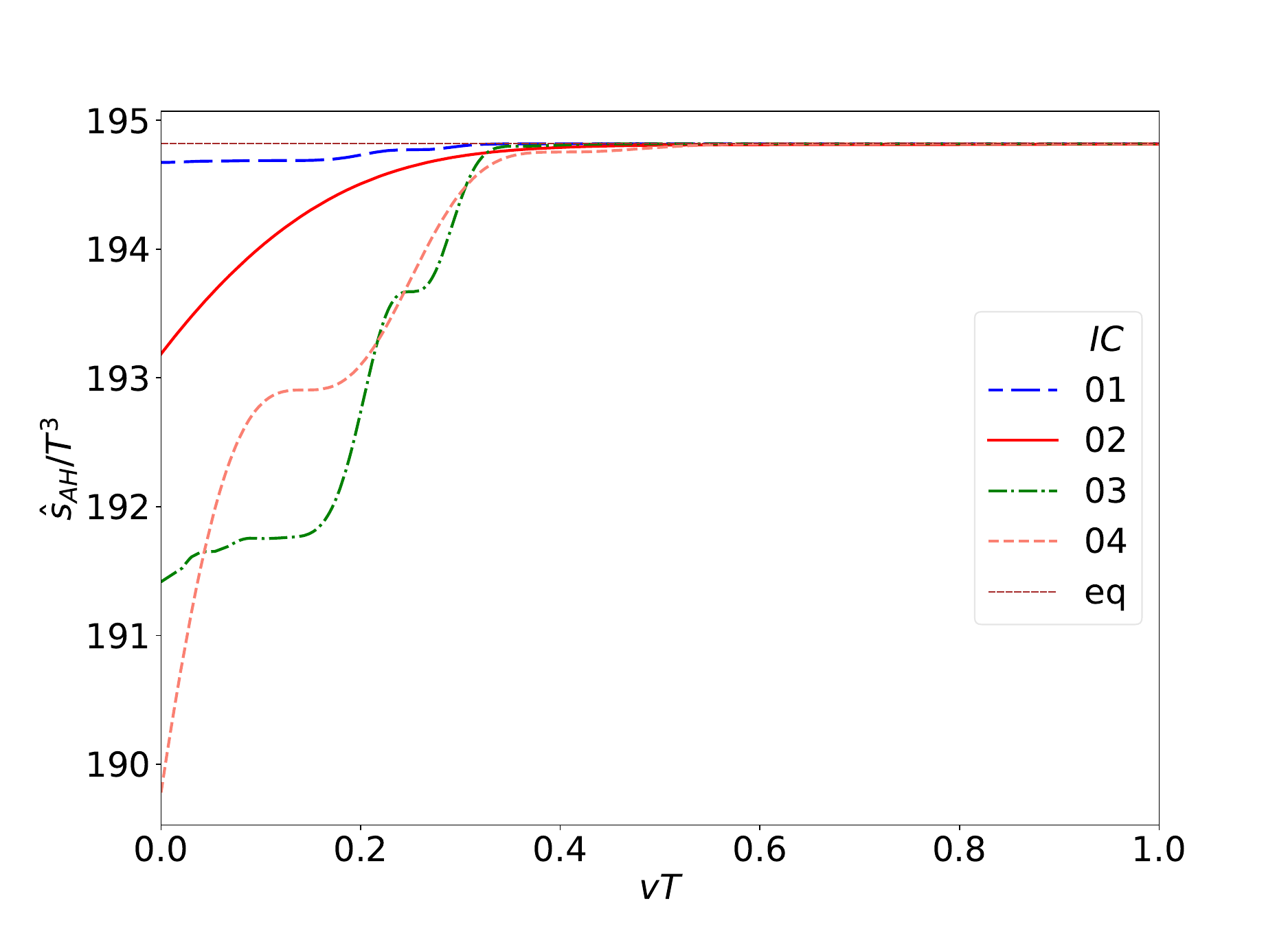}}
\subfigure[]{\includegraphics[width=0.5\textwidth]{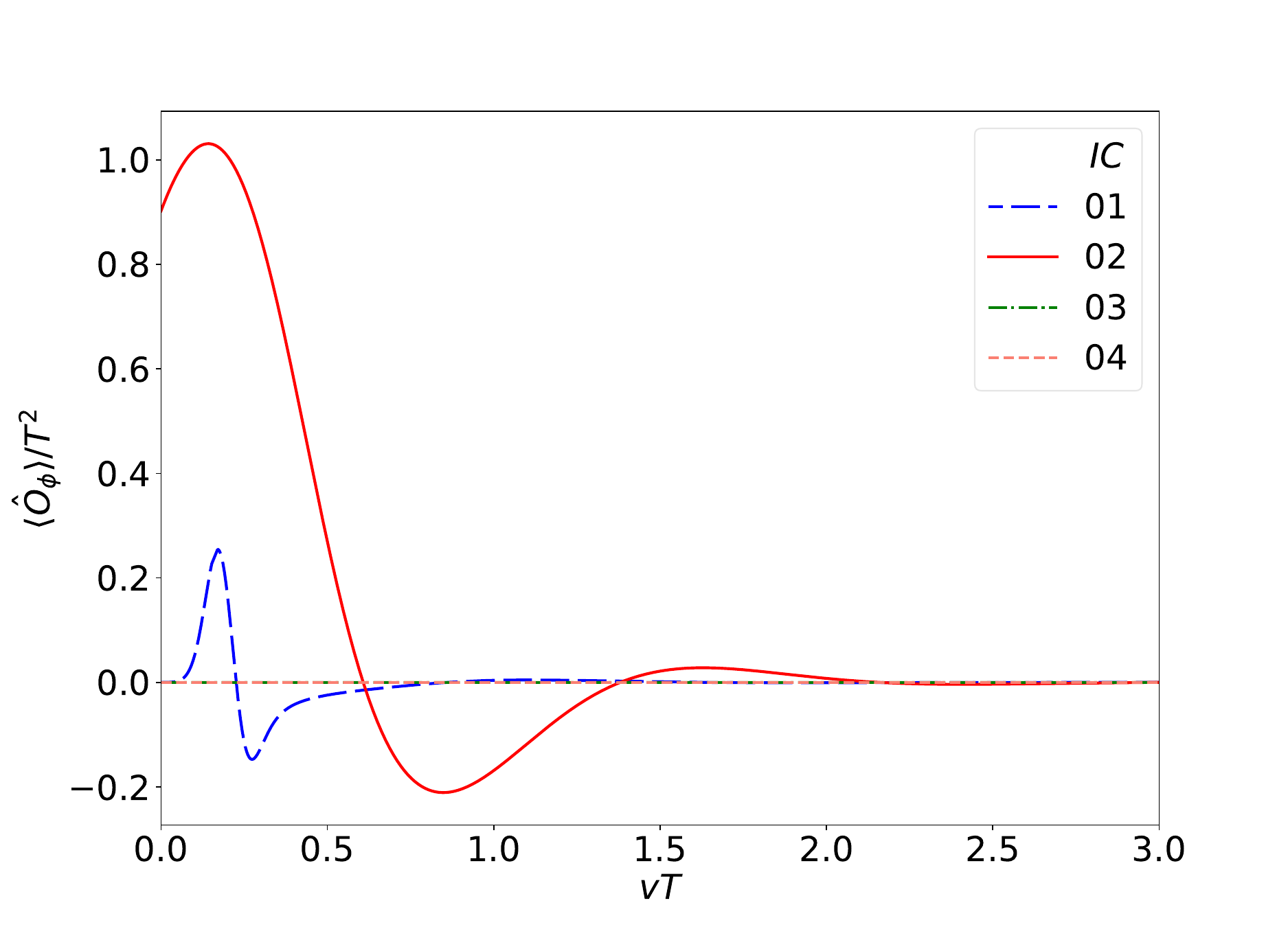}}
\caption{Numerical evolution of dimensionless ratios involving (a) the pressure anisotropy, (b) the non-equilibrium entropy density, and (c) the scalar condensate, all of them plotted for $\mu/T = 0$. The initial conditions (ICs) are given in Table \ref{tabICs}. In Fig. (a) for the pressure anisotropy, the salmon regions delimit the regions with WEC violation (which automatically imply DEC violation), while the gray region delimits the region with only DEC violation. In Fig. (b) the thin straight line gives the equilibrium value of the entropy density attained by the system at long times.}
\label{fig:muT0}
\end{figure*}

\begin{figure*}
\center
\subfigure[]{\includegraphics[width=0.49\textwidth]{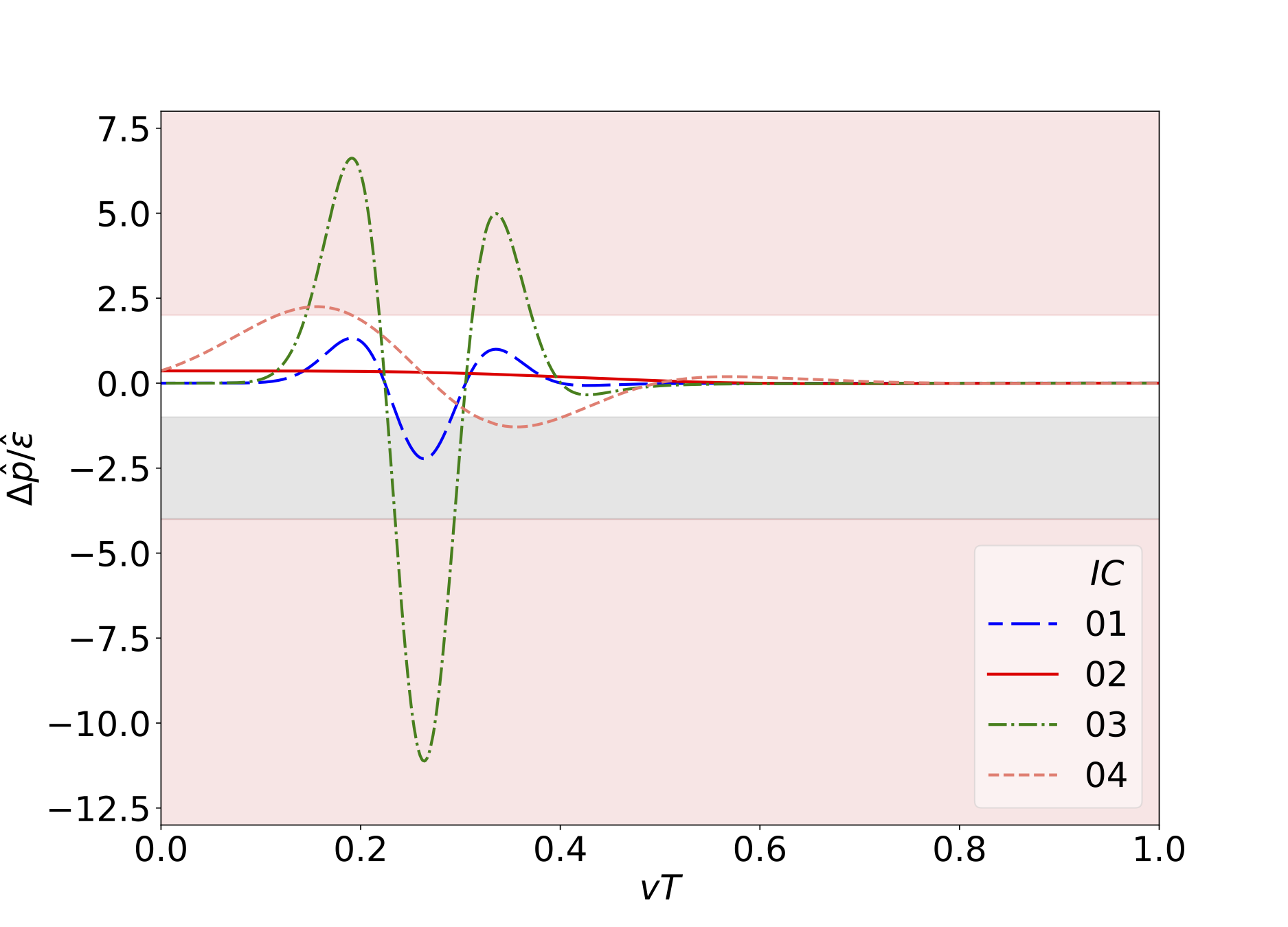}}
\subfigure[]{\includegraphics[width=0.49\textwidth]{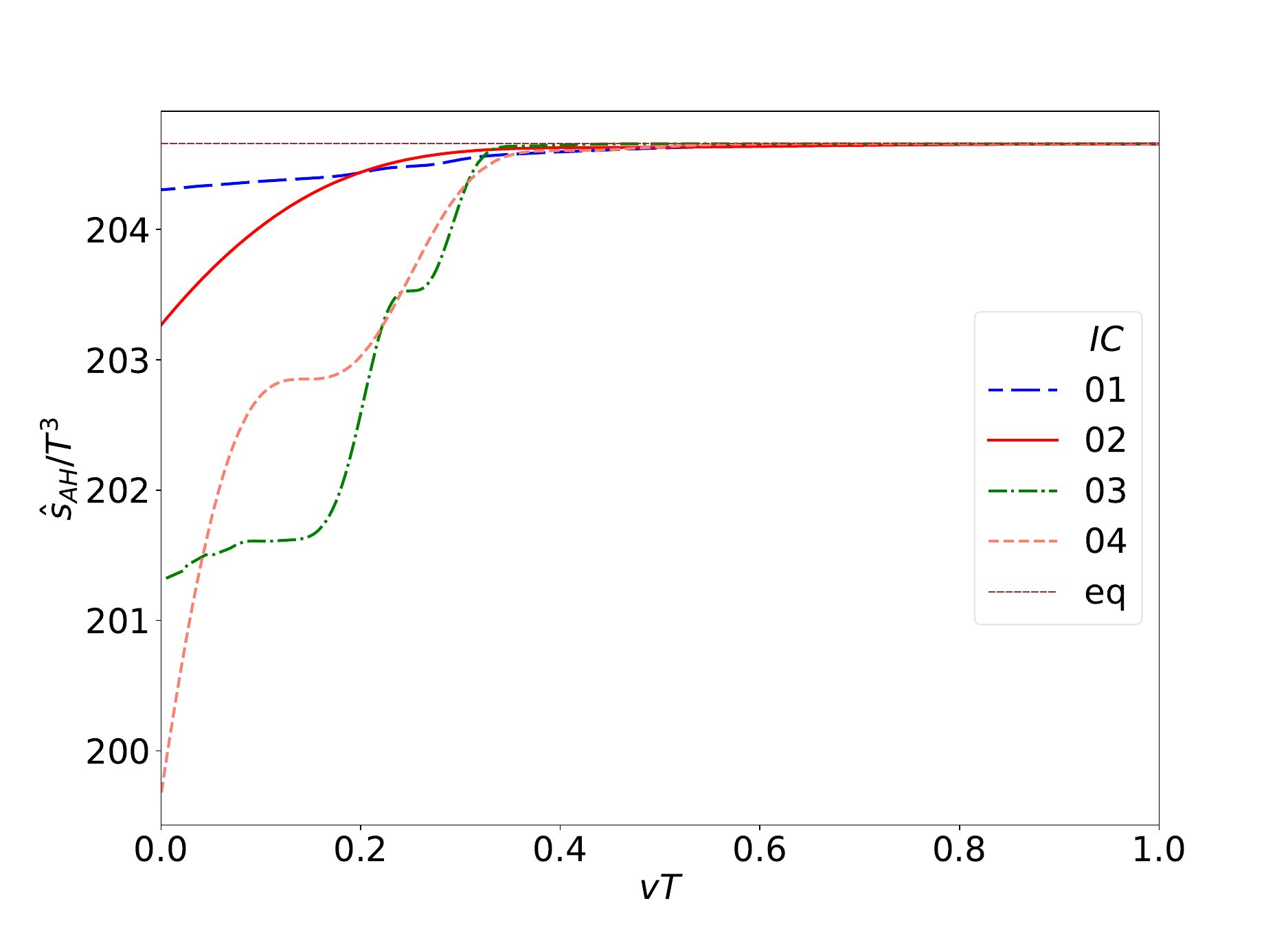}}
\subfigure[]{\includegraphics[width=0.5\textwidth]{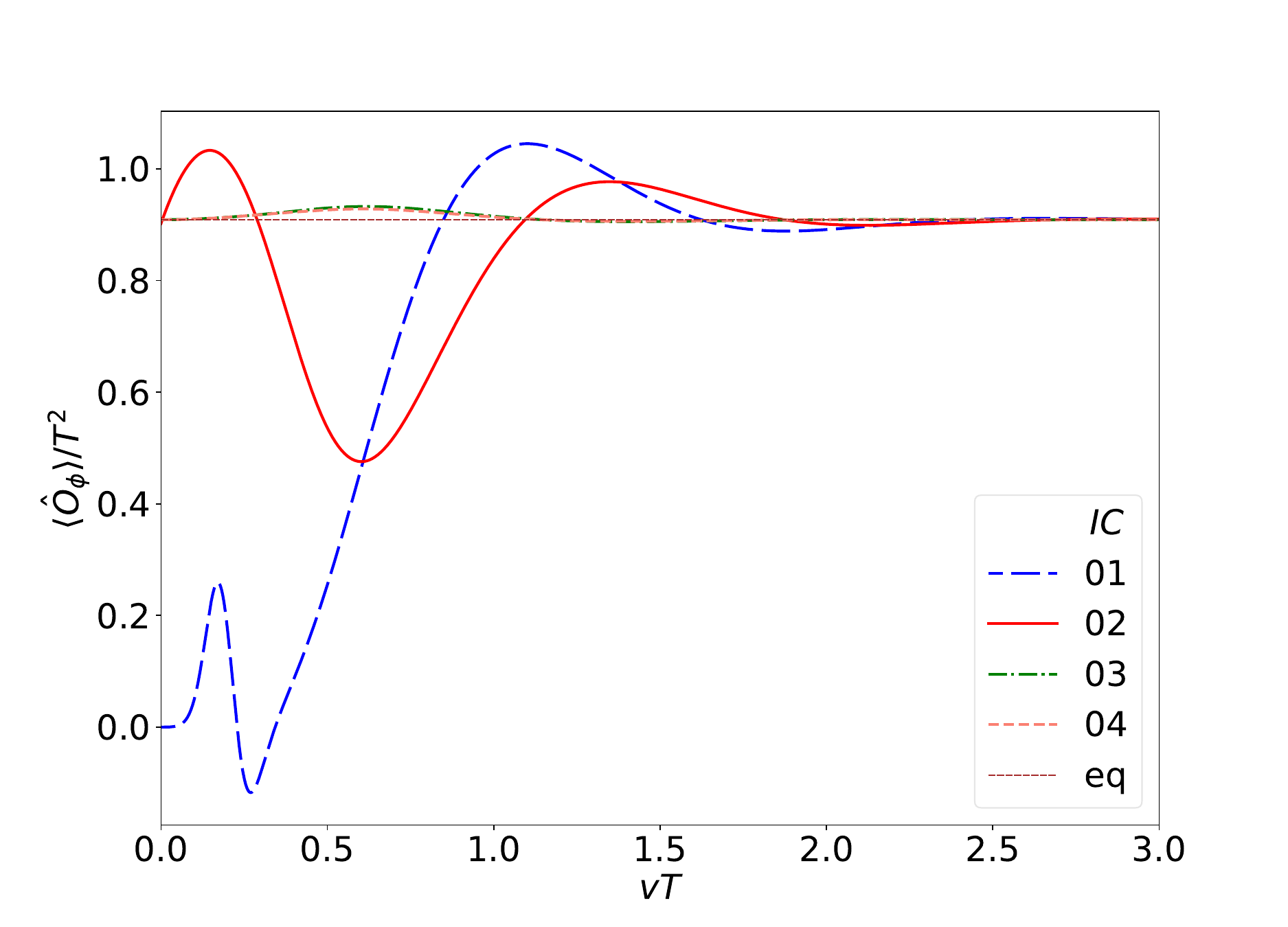}}
\caption{Numerical evolution of dimensionless ratios involving (a) the pressure anisotropy, (b) the non-equilibrium entropy density, and (c) the scalar condensate, all of them plotted for $\mu/T = 1$. The initial conditions (ICs) are given in Table \ref{tabICs}. In Fig. (a) for the pressure anisotropy, the salmon regions delimit the regions with WEC violation (which automatically imply DEC violation), while the gray region delimits the region with only DEC violation. In Figs. (b) and (c) the thin dashed lines give, respectively, the equilibrium value of the entropy density and of the scalar condensate attained by the system at long times.}
\label{fig:muT1}
\end{figure*}

\begin{figure*}
\center
\subfigure[]{\includegraphics[width=0.49\textwidth]{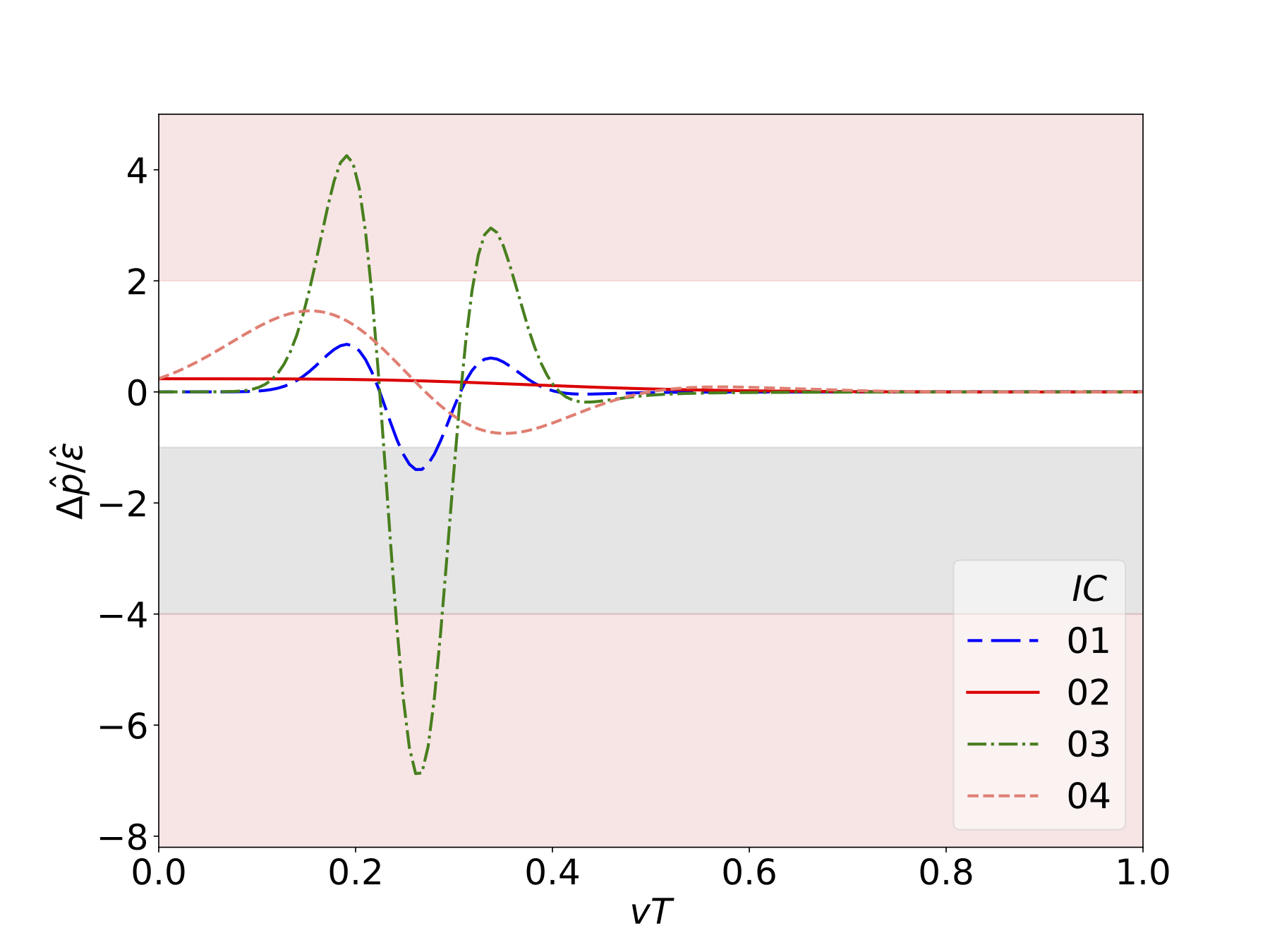}}
\subfigure[]{\includegraphics[width=0.49\textwidth]{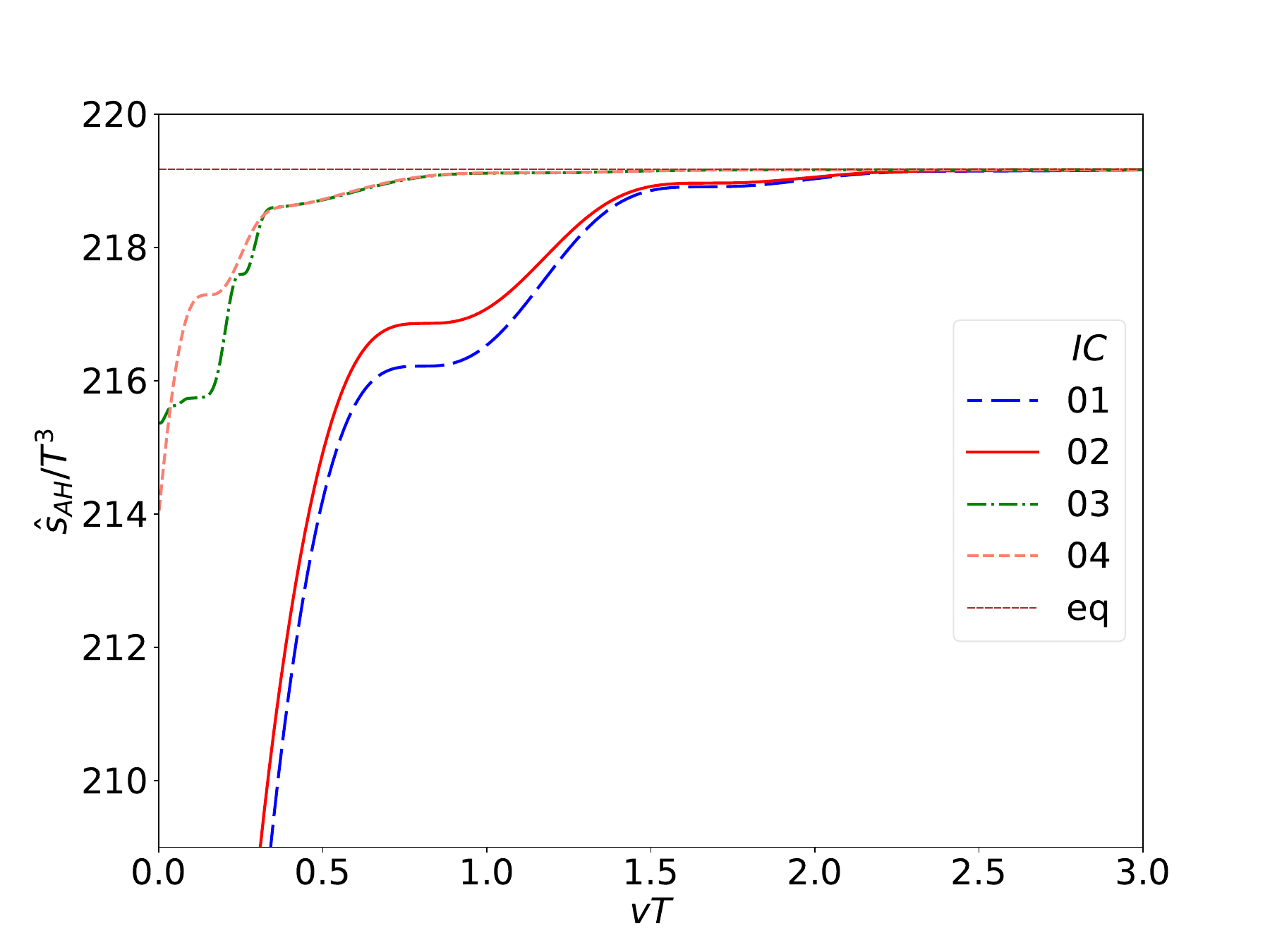}}
\subfigure[]{\includegraphics[width=0.5\textwidth]{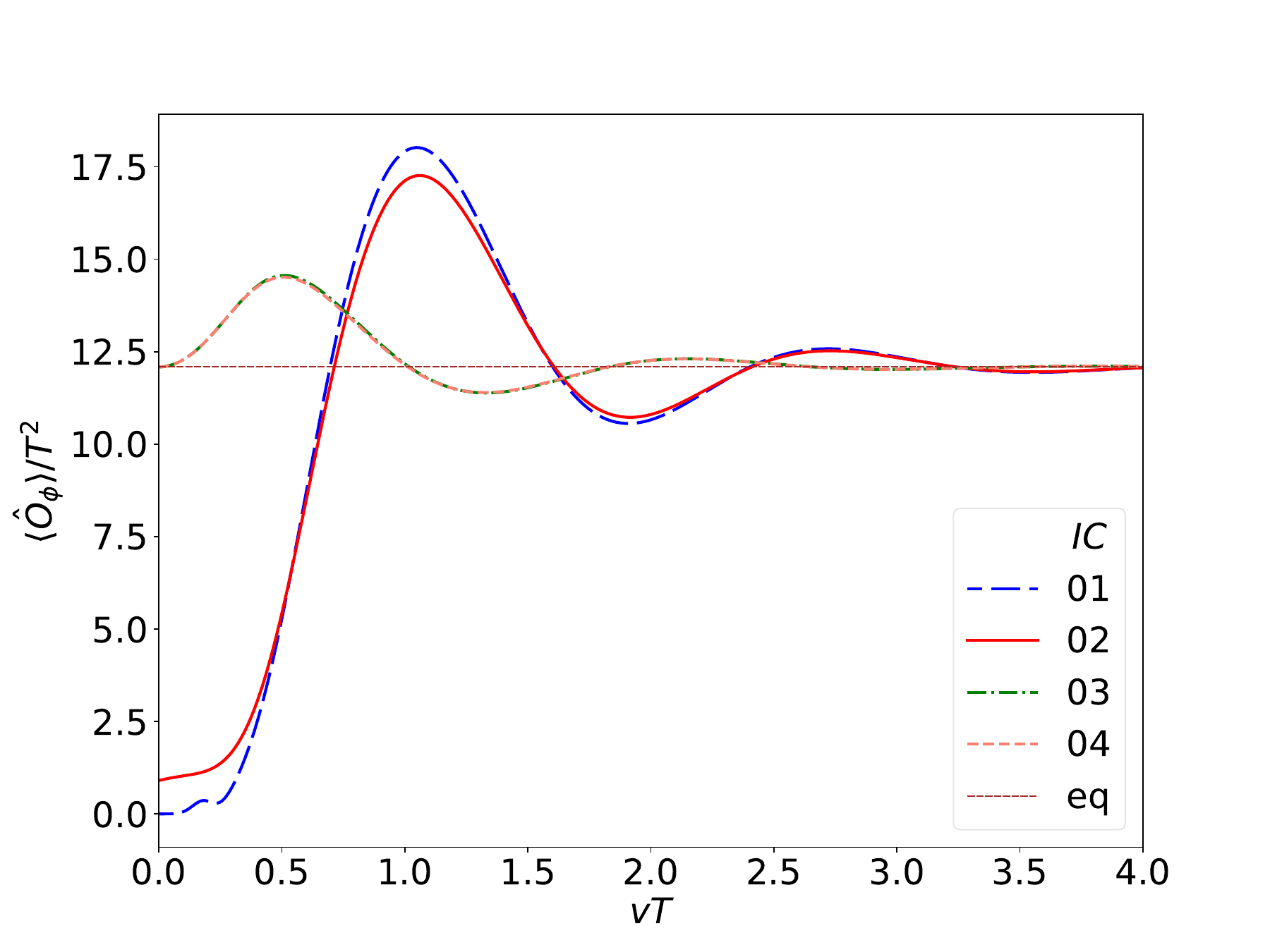}}
\caption{Numerical evolution of dimensionless ratios involving (a) the pressure anisotropy, (b) the non-equilibrium entropy density, and (c) the scalar condensate, all of them plotted for $\mu/T = \pi/\sqrt{2}$ (critical point). The initial conditions (ICs) are given in Table \ref{tabICs}. In Fig. (a) for the pressure anisotropy, the salmon regions delimit the regions with WEC violation (which automatically imply DEC violation), while the gray region delimits the region with only DEC violation. In Figs. (b) and (c) the thin dashed lines give, respectively, the equilibrium value of the entropy density and of the scalar condensate attained by the system at long times.}
\label{fig:muTcp}
\end{figure*}

\begin{figure*}
\center
\subfigure[]{\includegraphics[width=0.49\textwidth]{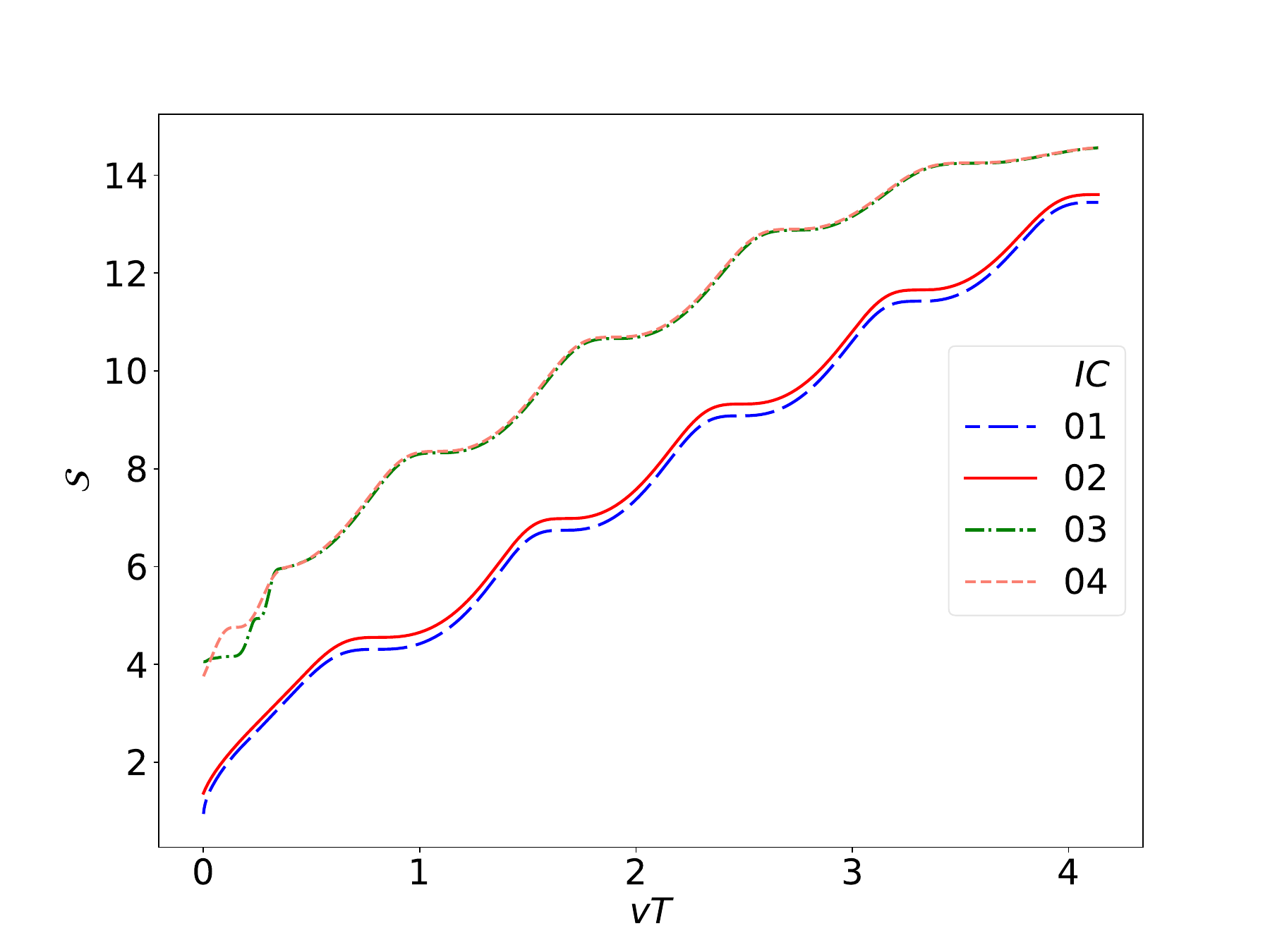}}
\subfigure[]{\includegraphics[width=0.49\textwidth]{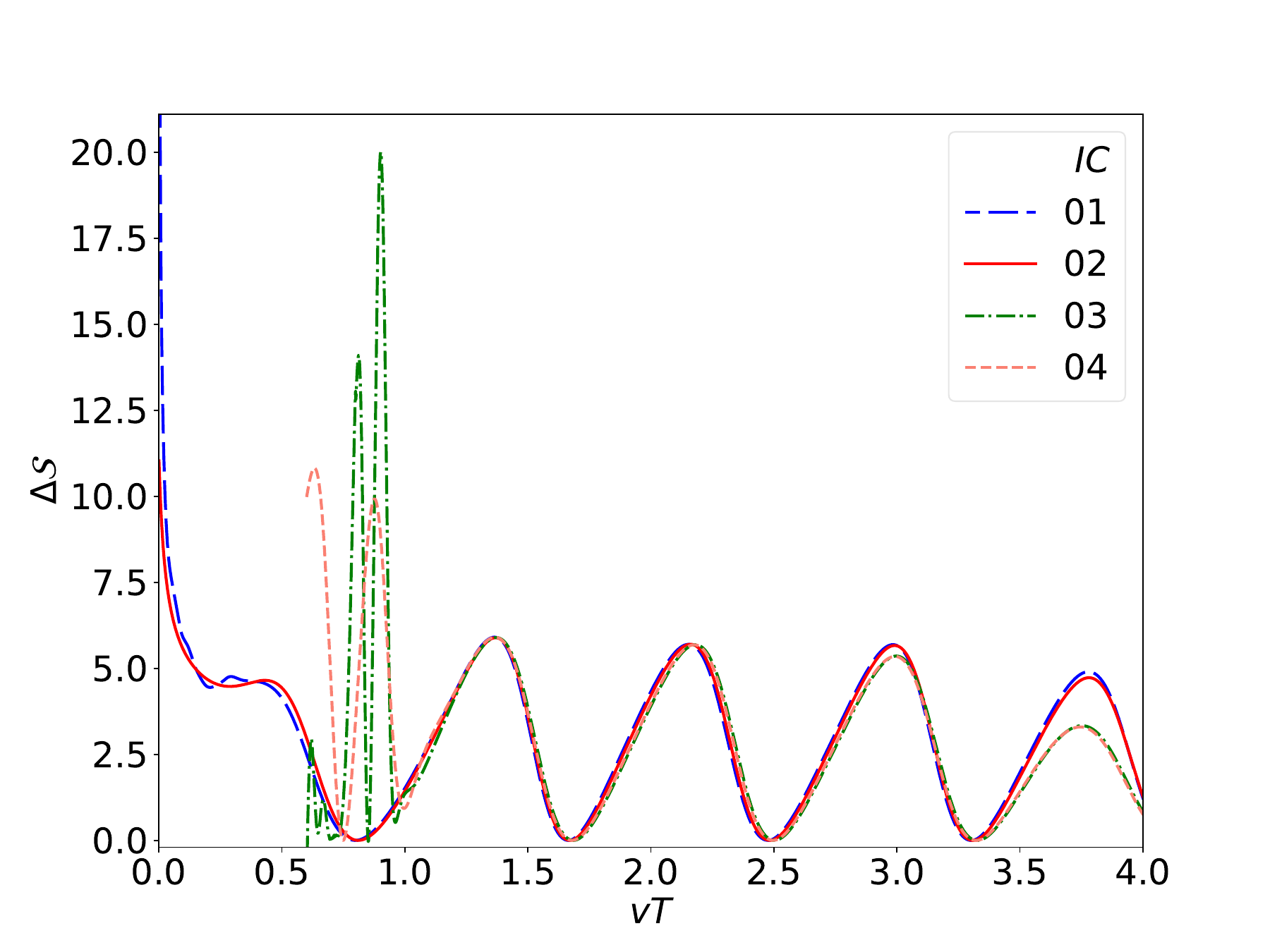}}
\subfigure[]{\includegraphics[width=0.49\textwidth]{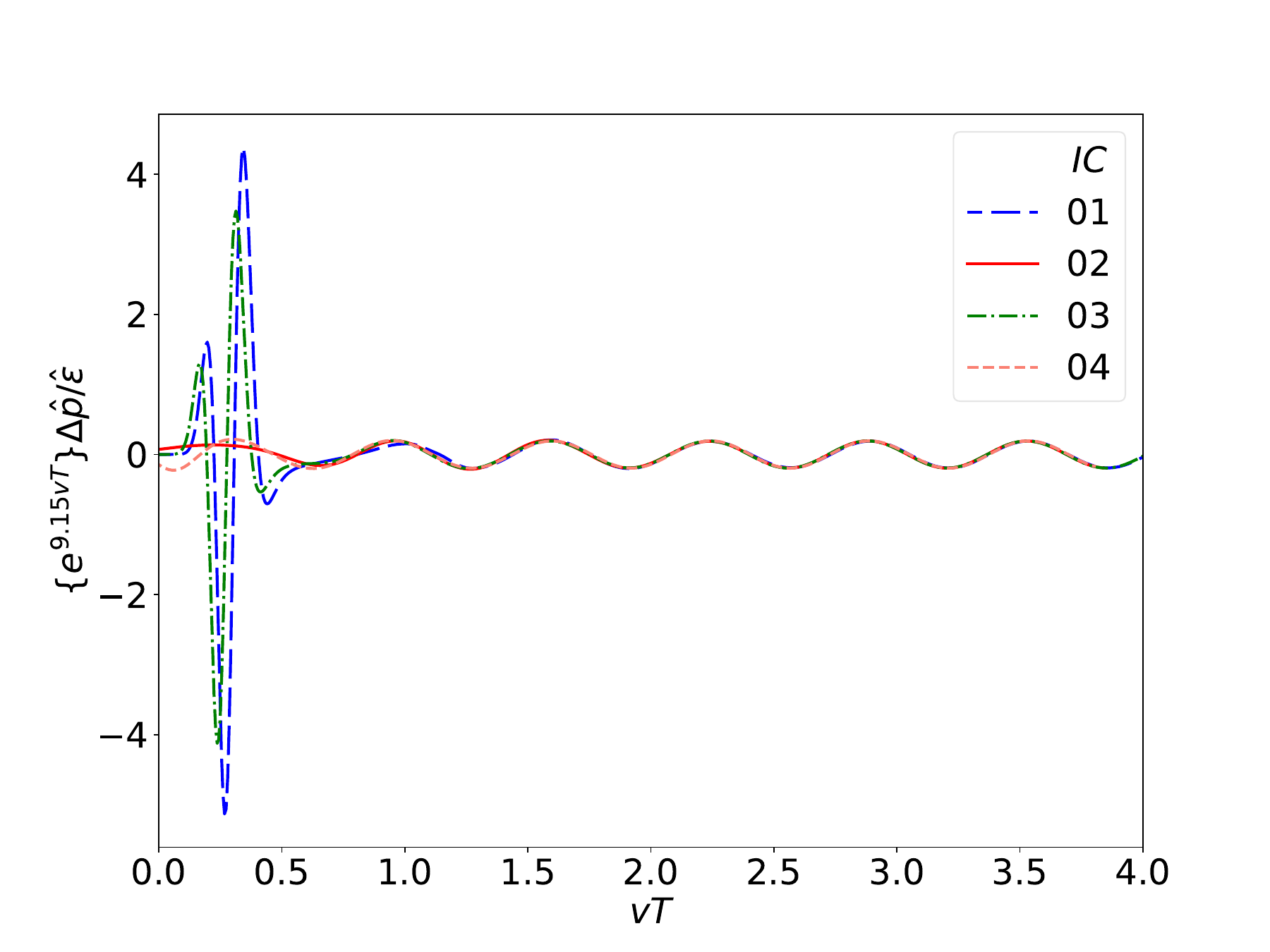}}
\subfigure[]{\includegraphics[width=0.49\textwidth]{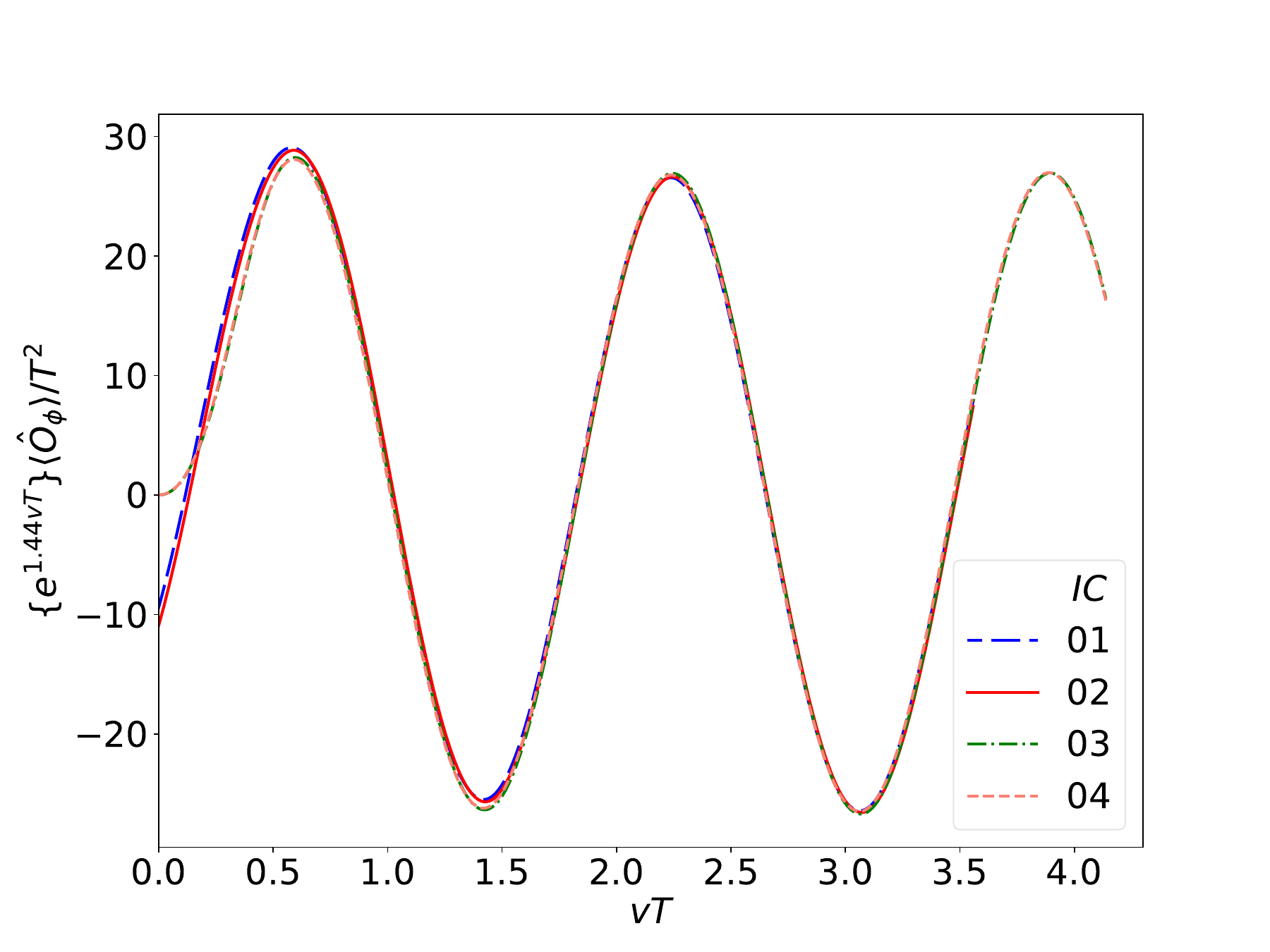}}
\caption{Time evolution of: (a) the logarithmic function $\mathcal{S}$ defined in Eq. \eqref{eq:logs}; (b) its first finite difference $\Delta\mathcal{S}$ (with the curves shifted to visually overlap); (c) the harmonic part of the pressure anisotropy (also with the curves shifted to visually overlap); (d) the harmonic part of the scalar condensate (also with the curves shifted to visually overlap). The initial conditions (ICs) are given in Table \ref{tabICs} and are plotted here for $\mu/T=\pi/\sqrt{2}$ (critical point).}
\label{fig:main}
\end{figure*}

In Figs. \ref{fig:muT0}, \ref{fig:muT1}, and \ref{fig:muTcp} we display the numerical time evolution for the different physical observables considering the initial data discussed in the previous section. One can notice that typically the scalar condensate only approaches its asymptotic equilibrium value for each $\mu/T$ at considerably larger times than the pressure anisotropy and the entropy density. Furthermore, for some initial data preserving all the energy conditions discussed before, one can observe transient violations of the DEC and even of the WEC when the fluid is still far from thermodynamic equilibrium. This has been also observed before for the 1RCBH model undergoing Bjorken flow \cite{Rougemont:2022piu} (and, as a particular case, also in the Bjorken flow of the SYM plasma \cite{Rougemont:2021qyk,Rougemont:2021gjm}), however, differently than in Bjorken flow, for the homogeneous isotropization dynamics of the 1RCBH model such energy condition violations are generally reduced as one increases the value of $\mu/T$.

Also as in the Bjorken flow, we observe for some initial data the formation of transient plateaus with zero entropy production when the system is still far from thermodynamic equilibrium. However, the interplay of these far from equilibrium plateaus and transient violations of energy conditions measured by the behavior of the pressure anisotropy is more complicated in the homogeneous isotropization dynamics. In fact, not all far from equilibrium plateaus in the entropy density are anticipating a posterior violation of energy conditions in the pressure anisotropy in the case of the homogeneous isotropization dynamics.

Moreover, from the analysis of several time evolutions of different initial conditions, for which figures \ref{fig:muT0} -- \ref{fig:muTcp} are quite representative, we observed for any considered value of $\mu/T$ the formation of multiple plateaus in the form of a stairway close to the respective equilibrium value for the entropy density. The plateaus produced in this late time near-equilibrium regime have no relation with the far-from-equilibrium plateaus which may or may not be produced in the time evolution of the system, depending on the initial conditions considered. Moreover, there are no violations of energy conditions near thermodynamic equilibrium, where the stairway structure of near-equilibrium plateaus for the entropy is observed. The near-equilibrium stairway structure for the entropy could easily have gone unnoticed due to the very close proximity of such plateaus, which requires a high numerical precision to be resolved. To clearly display this peculiar behavior we constructed the following logarithmic function with the non-equilibrium entropy density $\hat{s}_\textrm{AH}(vT;\mu/T)$ and its asymptotic equilibrium value $\hat{s}_\textrm{eq}(\mu/T)$,
\begin{equation}
\mathcal{S}(vT;\mu/T) = \ln\left\{ \frac{\hat{s}_\textrm{eq}(\mu/T)}{\hat{s}_\textrm{eq}(\mu/T)-\hat{s}_\textrm{AH}(vT;\mu/T)} \right\}.
\label{eq:logs}
\end{equation}
This led us to realize that the stairway to equilibrium entropy comprises a periodic formation of ever increasing plateaus asymptotically approaching its ceiling value.
In order to determine such a period at the critical point of the model as displayed in Fig. \ref{fig:main}, we calculated the finite difference of Eq. \eqref{eq:logs} to establish where the minima and maxima of $\mathcal{S}(vT;\mu/T)$ were. We then made use of the well known software {\sc harminv} \cite{harminv} to evaluate the mean value of the frequency of such extrema by means of the calculation of the second finite difference of Eq. (\ref{eq:logs}), which gave us, at the critical point (CP),\footnote{We remark that the quoted uncertainties refer primarily to the calculation of the frequency from the given numerical data. There are also uncertainties stemming from the numerical solutions of the EMD equations of motion which are not being accounted for here.}
\begin{align}
\frac{\nu_{\cal{S}}^{CP}}{T}=(1.21235 \pm 0.00002)\,\,\Rightarrow\,\, \tau_{\cal{S}}^{CP}T \approx (0.82484 \pm 0.00001).
\label{eq:nuScp}
\end{align}
It is important to clearly state that in Fig. \ref{fig:main} (a) the quantity $\mathcal{S}$ defined on Eq. \eqref{eq:logs} is not saturating at large times. In fact, from the given definition this quantity asymptotically diverges at asymptotic times, as the entropy density tends towards its equilibrium value. Of course, our numerical simulations cannot proceed to arbitrarily large times, so that the plot represents the results obtained up to the end time chosen for our simulations. If one wants to increase the end time of the simulations, in order to resolve the structure of the stairway to equilibrium entropy at larger times, due to the progressively smaller differences in magnitude with respect to the ceiling asymptotic value of equilibrium, one also needs to increase the numerical precision of the calculations, which will considerably augment the computation time of the simulations.

The periodicity of the oscillations of the pressure anisotropy near equilibrium is given by the real part of the lowest quasinormal frequency of the $SO(3)$ quintuplet channel of the 1RCBH model \cite{Critelli:2017euk,Finazzo:2016psx}. On the other hand, the exponential decay of such oscillations is associated to the imaginary part of that complex quasinormal frequency. In Fig. \ref{fig:main} (c) we removed the aforementioned exponential damping and employed the {\sc harminv} software to obtain the harmonic frequency of the pressure anisotropy oscillations near equilibrium evaluated at the critical point,
\begin{align}
\frac{\mathrm{Re}\left[\nu_{p}^{CP}\right]}{T}=(1.54887 \pm 0.00002)\,\,\Rightarrow\,\, \mathrm{Re}\left[\tau_{p}^{CP}\right]T \approx (0.645632 \pm 0.000008).
\label{eq:nuPcp}
\end{align}
The above result agrees with the real part of the lowest quasinormal frequency of the $SO(3)$ quintuplet channel of the 1RCBH model at the critical point as calculated in \cite{Critelli:2017euk}.\footnote{See Fig.\ 22 (a) of \cite{Critelli:2017euk} taking into account that $\omega=2\pi\nu$. Notice also that the coefficient of the argument in the time exponential used in Fig.\ \ref{fig:main} (c) is given by minus the imaginary part of the quasinormal frequency in Fig.\ 22 (b) of \cite{Critelli:2017euk}.}

Furthermore, the periodicity of the oscillations of the scalar condensate near equilibrium is given by the real part of the lowest quasinormal frequency of the $SO(3)$ singlet channel of the 1RCBH model \cite{Critelli:2017euk}. On the other hand, the exponential damping of such oscillations is associated to the imaginary part of that complex quasinormal frequency. In Fig. \ref{fig:main} (d) we removed the aforementioned exponential decay and used the {\sc harminv} software to obtain the harmonic frequency of the scalar condensate oscillations near equilibrium evaluated at the critical point,
\begin{align}
\frac{\mathrm{Re}\left[\nu_{\langle\hat{O}\rangle}^{CP}\right]}{T}=(0.60804 \pm 0.00005)\,\,\Rightarrow\,\, \mathrm{Re}\left[\tau_{\langle\hat{O}\rangle}^{CP}\right]T \approx (1.6446 \pm 0.0001).
\label{eq:nuOcp}
\end{align}
The above result agrees with the real part of the lowest quasinormal frequency of the $SO(3)$ singlet channel of the 1RCBH model at the critical point as calculated in \cite{Critelli:2017euk}.\footnote{See Fig.\ 24 (a) of \cite{Critelli:2017euk} taking into account that $\omega=2\pi\nu$. Notice also that the coefficient of the argument in the time exponential used in Fig.\ \ref{fig:main} (d) is given by minus the imaginary part of the quasinormal frequency in Fig.\ 24 (b) of \cite{Critelli:2017euk}.}

In Fig.\ \ref{fig:main} we show, for the four initial conditions in Table \ref{tabICs} evaluated at the critical point $\mu/T=\pi/\sqrt{2}$, the logarithmic function $\mathcal{S}$ defined in Eq. \eqref{eq:logs}, its first finite difference $\Delta\mathcal{S}$ with each curve shifted to visually overlap for presentation purposes of our main results, and the harmonic parts of the pressure anisotropy and of the scalar condensate oscillations, also shifted to visually overlap. From the results in Eqs. \eqref{eq:nuScp}, \eqref{eq:nuPcp} and \eqref{eq:nuOcp} one sees that at the critical point the period of plateau formation in the stairway to equilibrium entropy is $\sim 28\%$ larger than the period of harmonic oscillations of the pressure anisotropy close to thermodynamic equilibrium, while being $\sim 50\%$ smaller than the period of harmonic oscillations of the scalar condensate in the linear regime.

Interestingly, we repeated the above calculations also for $\mu/T = \{0,1,2\}$ and found that in all cases the period of plateau formation in the stairway to equilibrium entropy is approximately half the period of oscillations of the slowest quasinormal mode of the system. That is always the lowest quasinormal mode associated to the scalar condensate, except for the particular case of the purely thermal SYM plasma defined at $\mu/T = 0$ with zero scalar condensate, and in such a case the slowest quasinormal mode is associated to the pressure anisotropy of the medium. The results are summarized in Table \ref{tabQNMs}.

\begin{table}[h]
\centering
\begin{tabular}{|c||c|c|c|}
\hline
$\mu/T$ & $\tau_\mathcal{S} T$ & $\textrm{Re}\left[\tau_p\right] T$ & $\textrm{Re}\left[\tau_{\langle\hat{O}\rangle}\right] T$ \\
\hline
\hline
0 (SYM) & 0.320(8) & 0.641140(7) & \textbf{---} \\
\hline
1 & 0.734(2) & \, 0.6360716(2) & \, 1.54841(2) \\
\hline
2 & \;\;\, 0.76836(2) & \;\;\;\; 0.632959465(4) &\;\;\;\; 1.5501881(2)  \\
\hline
$\pi/\sqrt{2}$ (CP) & \;\;\, 0.82484(1)&0.645632(8) & 1.6446(1) \\
\hline
\end{tabular}
\caption{Dimensionless periods for the different physical observables near thermodynamic equilibrium.}
\label{tabQNMs}
\end{table}

\section{Conclusions and Perspectives}
\label{sec:conc}

In the present work, we analyzed the homogeneous isotropization dynamics of the top-down 1RCBH holographic model, including for the first time the calculation of the time evolution of its non-equilibrium entropy density. We found that generally the scalar condensate takes a considerably longer time than the pressure anisotropy and the entropy density to approach the respective equilibrium values. Furthermore, for some initial data preserving all the energy conditions, transient violations of the dominant and even of the weak energy conditions are observed when the fluid is still far from thermodynamic equilibrium, with the magnitude of such violations getting reduced as the chemical potential of the medium is increased (contrary to what happens in the Bjorken flow of the same model).

Moreover, a new feature disclosed in the present work is the formation of a stairway to equilibrium entropy at late times. This stairway is observed for all the initial data analyzed undergoing homogeneous isotropization dynamics in the 1RCBH model and comprises the formation of a periodic sequence of several close plateaus in the entropy density near thermodynamic equilibrium. We also found that the period of plateau formation in this near-equilibrium stairway structure for the entropy is always half the period of oscillations of the slowest quasinormal mode of the system. For finite density states of the present model, the slowest quasinormal mode is always associated to the late time equilibration of the scalar condensate. In the particular case of the purely thermal SYM plasma defined at zero density and vanishing scalar condensate, the slowest quasinormal mode of the system is associated to the late time equilibration of the pressure anisotropy.

At this point, we believe we may provide a more general perspective about the mains results obtained in the present work, concerning the homogeneous isotropization dynamics of the 1RCBH model (having the purely thermal SYM plasma as a particular case), and also some of the main results obtained in previous works concerning the Bjorken flow dynamics of the same model \cite{Rougemont:2021qyk,Rougemont:2021gjm,Rougemont:2022piu}. In fact, one may ask e.g. whether some qualitative results reported in those different works are model-dependent and/or dynamics-dependent.

We begin by addressing model dependence. Since the 1RCBH model analyzed in the present work reduces to the purely thermal SYM plasma at zero density and vanishing scalar condensate, and since several other holographic models at finite density also reduce to the purely thermal SYM plasma in the zero chemical potential limit, we expect that some features observed both in the 1RCBH model at finite and zero density, will be also generally displayed by other holographic models. Such an expectation includes the formation of a stairway structure for the entropy near thermodynamic equilibrium in the homogeneous isotropization dynamics, which is something that can be tested in other models.

We now discuss dynamics-dependence, i.e. the dependence of some qualitative results on the specific kind of far from equilibrium dynamics considered. The stairway structure for the entropy near thermodynamic equilibrium disclosed in the present work for the homogeneous isotropization dynamics has not been observed in the Bjorken flow dynamics. In fact, as aforementioned, the formation of plateaus for the entropy near equilibrium in the homogeneous isotropization dynamics is periodic, with its period corresponding to half the period of oscillations of the observable which takes longer to equilibrate in the system. Consequently, in the homogeneous isotropization dynamics, the period of oscillations of the slowest observable to equilibrate in the system is strongly correlated with the dissipative dynamics of the system associated to the irreversible production of entropy near thermodynamic equilibrium, which turns out to be also periodic. Very interestingly, since the entropy cannot oscillate without violating the second law of thermodynamics, it finds a way of developing a periodicity by creating a stairway structure with periodic formation of plateaus. On the other hand, in the Bjorken flow dynamics, it was found that the pressure anisotropy and the scalar condensate hydrodynamize at late times by going into analytical curves which decrease towards their respective asymptotic values without oscillating \cite{Rougemont:2021qyk,Rougemont:2021gjm,Rougemont:2022piu}. Correspondingly, we have also not observed the formation of a stairway structure for the entropy in the late time evolution of the system in the Bjorken flow dynamics.\footnote{For the particular case of the purely thermal SYM plasma, analytical expressions for the late-time hydrodynamic curves for the entropy are known, from which it is clear that a stairway structure with periodic plateau formation for the near-equilibrium entropy cannot be produced in the Bjorken flow dynamics --- see e.g. Eqs. (46) of \cite{Rougemont:2021gjm} and Eq. (23b) of \cite{Chesler:2009cy}.} Moreover, one of the main conclusions we previously reached for the Bjorken flow dynamics, regarding the statement that exact plateaus for the entropy when the system is still far-from equilibrium always anticipates posterior violations of the dominant energy condition from below, is something that we have found in the present work to not hold for the homogeneous isotropization dynamics. The present analysis provides, therefore, a broader picture regarding some dynamics-dependent features of far-from-equilibrium strongly coupled quantum fluids described by holographic models.

More importantly, the present analysis also indicates that one may \emph{predict} some general features of the entropy or of the pressure anisotropy and of the scalar condensate by just knowing the behavior of some other observables. Indeed, by detecting the presence or absence of a stairway structure for the entropy near thermodynamic equilibrium, from the analysis of both, the homogeneous isotropization dynamics and the Bjorken flow dynamics, it is clear that one could correctly anticipate whether the pressure anisotropy and the scalar condensate tend towards their asymptotic values with or without oscillating around them, respectively (and vice-versa). Not only that, but if the stairway structure is detected, one can further \emph{predict} the period of oscillations near thermodynamic equilibrium of the observable which takes longer to equilibrate as being twice the period of plateau formation in the stairway (and vice-versa).

As a future perspective, it would be interesting to investigate some important points related specifically to the Bjorken flow dynamics. In particular, the relation between transients of the Bjorken flow and the homogeneous quasinormal modes \cite{Janik:2006gp,Banerjee:2022aub,Heller:2013fn}, and the relation disclosed here between the homogeneous quasinormal modes and the periodic stairway structure for the entropy density near thermodynamic equilibrium in the homogeneous isotropization dynamics, may suggest that those facts may be somehow connected.

Moreover, as another perspective for future works, possibly the implementation of machine learning techniques can further help in the task of recognizing non-obvious patterns and correlations between different physical observables, as in the recent work of Ref. \cite{Jejjala:2023zxw}. For instance, in that work, a deep neural network applied to a different holographic model allowed for the reconstruction of the boundary non-equilibrium entropy from data concerning some characteristic features of the boundary pressure anisotropy.

As a longer-term perspective, it would be interesting to investigate entropy production and thermalization in bottom-up holographic EMD models tailored for a quantitative description of the quark-gluon plasma produced in heavy-ion collisions \cite{MUSES:2023hyz}, as in the constructions discussed in \cite{Rougemont:2023gfz,Hippert:2023bel}.

\acknowledgments
We thank Lorenzo Gavassino for insightful questions and comments regarding some of our results. We acknowledge financial support by National Council for Scientific and Technological Development (CNPq) under grant number 407162/2023-2. W.B. is grateful to S\~{a}o Paulo Research Foundation (FAPESP) for the received support under grant number 2022/02503-9.




\bibliographystyle{apsrev4-2}
\bibliography{bibliography,extrabiblio} 

\end{document}